\documentclass[reprint, onecolumn, secnumarabic, amssymb, nobibnotes, aps, prfluids]{revtex4-2}
\usepackage{graphicx}
\usepackage{epstopdf, epsfig}
\usepackage{amsmath}

\usepackage{multirow}
\usepackage[dvipsnames]{xcolor}
\usepackage{cancel}
\usepackage{lineno}
\usepackage{subcaption}
\usepackage{caption}
\usepackage{url}
\DeclareRobustCommand{\CORR}[1]{{\color{black}{#1}}}

\newcommand{\R}{\mathcal{R}}

\providecommand\bnabla{\boldsymbol{\nabla}}
\providecommand\bcdot{\boldsymbol{\cdot}}

\begin{document}

\title{Galerkin reduced-order model for two-dimensional Rayleigh-Bénard convection}
\begin{abstract}
In this work, Galerkin projection is used to build reduced-order models (ROM) for two-dimensional Rayleigh-Bénard (RB) convection with no-slip walls. We compare an uncoupled projection approach that uses separate orthonormal bases for velocity and temperature with a coupled formalism where the equations are projected onto a single basis combining velocity and temperature components.
Orthonormal bases for modal projection are obtained as the eigenfunctions of the controllability Gramian of the linearized RB equations, eliminating the need for DNS snapshot databases required by traditional POD-based approaches.
Various coupled and uncoupled ROMs with different numbers of modes are generated and validated against direct numerical simulations (DNS) over a wide range of Rayleigh numbers. 
One of the objectives is to determine their domain of validity as a function of the system dimension and the Rayleigh number. DNS and ROM results are compared in terms of mean vertical profiles, heat flux, flow structures, dynamical regimes and energy spectra. 
Crucially, unlike previous POD-based Galerkin models for thermal convection, these ROMs do not require closure models and remain numerically stable. The coupled approach shows better agreement with DNS in terms of mean vertical profiles and Nusselt number scaling. The capabilities of  these models are exploited to conduct a detailed bifurcation analysis at $Pr = 10$ using Poincaré sections and Lyapunov exponents, precisely identifying the transitions between periodic, quasiperiodic, and chaotic states with significant reductions of computational cost.
\end{abstract}
\author{Enrique Flores-Montoya}
\email{efloresm.ca@gmail.com}
\affiliation{Aerothermals department, ITP Aero, 28108 Alcobendas, Spain}
\author{Andr\'e V. G. Cavalieri}
\email{andre@ita.br}
\affiliation{Divis\~ao de Engenharia Aeroespacial, Instituto Tecnol\'ogico de Aeron\'autica, S\~ao Jos\'e dos Campos, SP, Brazil}
\date{\today}
\maketitle

\section{Introduction}\label{sec:Intro}
Rayleigh-Bénard convection is present in various natural phenomena such as atmospheric and oceanic flows, as well as in technological processes such as crystal growth and heat transfer systems~\cite{bodenschatz2000recent, manneville2006rayleigh}. This canonical example of buoyancy-driven flow exhibits hallmark features of nonlinear dynamics, such as pattern formation and chaotic behavior~\cite{busse1967stability,busse1978non}. Its rich dynamical behavior together with the relative ease of producing accurate experimental data have motivated its study since the seminal works of Lord Rayleigh and Bénard~\cite{benard1901tourbillons, rayleigh1916lix}.

In Rayleigh-Bénard (RB) convection, a fluid layer bounded by thermally conductive walls is heated from below and cooled from above. Two non-dimensional parameters govern the dynamics of the problem: the Prandtl, $Pr$, and the Rayleigh number, $Ra$. The Prandtl number is the ratio between the kinematic viscosity, $\nu$, and the thermal diffusivity of the fluid, $\kappa$. The Rayleigh number, defined as $Ra=\sigma g h^3 \Delta T/\nu\kappa$, quantifies the balance between buoyant driving forces and dissipative effects. Here, $\sigma$ is the thermal expansion coefficient of the fluid, $g$ is the gravitational acceleration, $\Delta T$ is the imposed temperature difference between the walls and $h$ is the height of the channel. In experiments, the Rayleigh number can be varied by increasing the temperature difference across the fluid layer. When the Rayleigh number reaches a critical value, $Ra_c$, buoyancy forces destabilize the fluid and convection develops. The critical $Ra$ is independent of the Prandtl number and is approximately 1708 for no-slip boundary conditions and 658 for stress-free boundaries~\cite{busse2005transition, chandrasekhar2013hydrodynamic}. Upon the onset of convection, the heat flux across the fluid layer, generally quantified using the Nusselt number, $Nu$, increases with the Rayleigh number. Because of its rich dynamical behavior, RB convection has been the subject of numerous theoretical and numerical studies~\cite{moore1973two, yahata1982transition, yahata1983period, pallares1996natural, zienicke1998bifurcations, pallares1999flow}. When the $Ra$ number increases, periodic and quasiperiodic states typically precede the onset of chaos~\cite{mclaughlin1982transition, curry1984order, paul2012bifurcation}. However, the sequence of bifurcations and routes to chaos depends on the Prandtl number and the boundary conditions. In low Prandtl number fluids, convection shows a complex bifurcation scenario involving 3D flow patterns~\cite{paul2012bifurcation}. For large Prandtl number fluids, 2D rolls survive up to large Rayleigh numbers making 2D simulations relevant for this regime.  

The complexity and elevated computational cost of the Navier-Stokes equations have motivated the development of reduced-order models (ROM)~\cite{waleffe1997self, noack2003hierarchy, rowley2017model, brunton2022data}. ROMs simplify the analysis of physical mechanisms by reducing the problem to the interactions among a few coherent flow structures.
In Galerkin ROMs, the {partial differential equations} (PDE) that describe the dynamics of the system are projected onto a set of modes. This reduces the PDE system to a set of {ordinary differential equations} (ODE) that describes the evolution of the mode amplitude coefficients. In this kind of models, orthogonal bases are often preferred since they simplify the energy representation and make mode growth or decay easier to analyze~\cite{cavalieri2021structure}.
Galerkin ROMs also facilitate the application of well-established tools from control theory and offer promising applications for nonlinear flow analysis and control~\cite{brunton2022data, liao2025convex}. For these applications to be effective, ROMs must accurately capture several key aspects of the flow physics: dominant flow structures, integral quantities (total heat flux, friction coefficient), system dynamics (periodic, chaotic regimes), and spectral characteristics. The ability to reproduce these features enables ROMs to serve as efficient surrogates for full-order models in real-time control applications and state observers~\cite{rowley2017model}. This is particularly valuable for systems requiring rapid computation such as feedback control where full numerical simulation would be computationally prohibitive.

In RB convection, model reduction via Galerkin projection has been extensively used to analyze the system dynamics~\cite{saltzman1962finite, lorenz1963deterministic,  manneville1983two}.  In the seminal work of~\citet{saltzman1962finite}, RB equations with free-slip boundary conditions are  projected  onto an orthogonal basis of trigonometric functions, yielding a system of ODEs for the mode amplitudes. From this formulation,~\citet{lorenz1963deterministic} derived the well-known Lorenz dynamical system by truncating the expansion down to three degrees of freedom. This approach---projecting velocity and temperature onto separate  bases with independent amplitude coefficients---has been exploited in numerous subsequent works to analyze bifurcations and transition to chaos~\cite{manneville1983two, paul2011bifurcations, paul2012bifurcation}. These models are generally derived for free-slip boundary conditions, which are difficult to realize experimentally but permit the analytical treatment of the problem by expanding the solution into sine/cosine functions in the vertical direction.

Extending Galerkin methods to no-slip boundary conditions introduces some complexity in the generation of the modal basis. One approach is to use Galerkin spectral methods with a large number of modes as a discretization technique for DNS. In this spirit, Giralt and coworkers employed Galerkin spectral methods with $n=\mathcal{O}(10^4)$ degrees of freedom to analyze bifurcations and stability in RB convection within cubical cavities~\cite{puigjaner2004stability, puigjaner2006bifurcation, puigjaner2008bifurcation, puigjaner2011steady}. Their divergence-free basis functions, constructed from combinations of trigonometric and hyperbolic functions, satisfy no-slip conditions but require finding roots of nonlinear equations. While accurate, such high-dimensional systems do not constitute reduced-order models in the sense of providing computational savings or dynamical insight through truncation. 

True ROMs for no-slip RB convection require constructing low-dimensional bases that capture the essential dynamics. 
Recently, data-driven approaches based on Proper Orthogonal Decomposition (POD) have been applied to RB convection with no-slip walls~\cite{podvin2001low, podvin2012proper, soucasse2019proper}. \citet{podvin2001low} developed a Galerkin ROM from POD modes for a differentially heated cavity and found that low-dimensional models required closure modeling to prevent numerical blow-up. \citet{podvin2012proper} extended this approach to turbulent 3D Rayleigh-Bénard convection in a rectangular cavity at high Rayleigh numbers. In a recent work, \citet{soucasse2019proper} used POD modes from 3D DNS snapshots to build a Galerkin ROM for a buoyant cavity. The 7-mode ROM includes a closure model combining linear and cubic terms and was capable of reproducing the low frequency dynamics of the large-scale reorientations. A common feature of these POD-based approaches is the need for DNS snapshot databases to construct the modal basis and the requirement of empirically tuned closure models to ensure numerical stability. A distinguishing aspect of these works~\cite{podvin2001low, podvin2012proper, soucasse2019proper} is the use of coupled modal bases where each POD mode contains both velocity and temperature components scaled by a single amplitude coefficient. This coupled basis approach accounts for the physical coupling between momentum and temperature and requires balancing their energy contributions through a meaningful inner product.

An alternative to POD that avoids the need for snapshot data has recently emerged~\cite{cavalieri2021structure, cavalieri2022reduced, cavalieri2023non, mccormack2024multi}. \citet{cavalieri2022reduced} developed Galerkin ROMs for 3D plane Couette flow using an orthogonal modal basis obtained from the eigenvectors of the controllability Gramian of the linearized equations. 
Controllability modes correspond to the POD modes of the linear system forced with white noise, thereby representing the most energetic and controllable coherent flow structures of the linear system.
Crucially, the resulting ROMs do not require closure models to remain numerically stable---a major advantage over earlier reported POD-based approaches~\cite{ahmed2021closures}. 
Despite the considerable degree of truncation, these ROMs are able to reproduce the main turbulence statistics with reasonable accuracy at a very low computational cost.
This methodology has been extended to mixing layers~\cite{cavalieri2023non}, used to compute invariant solutions in plane Couette flow~\cite{mccormack2024multi}, and recently employed by~\citet{maia2025turbulence} to design control strategies that laminarize turbulent flow, demonstrating the utility of these models for flow control applications.
Beyond its scientific interest, RB convection plays a key role in technological applications such as Czochralski crystal growth~\cite{muller1988convection}, where convection-induced inhomogeneities in dopant distribution must be controlled~\cite{gunzburger2002controlling, alloui2018control}. Galerkin ROMs offer a promising platform for developing active control strategies, as explored by~\citet{munch2008galerkin}. However, the implementation and validation of quantitatively accurate ROMs with representative boundary conditions is a necessary prerequisite for successful control applications.

In this work, we develop reduced-order models for two-dimensional RB convection with no-slip walls via Galerkin projection using controllability modes. Two projection strategies are systematically compared. In the uncoupled approach, independent modal bases for temperature and velocity are used, following the classical Galerkin tradition.
 In the coupled approach, RB equations are projected onto a single orthogonal basis combining velocity and temperature components scaled by a common amplitude coefficient. For both strategies, the orthogonal bases are obtained numerically as eigenfunctions of the controllability Gramian of the linearized equations, enabling no-slip boundary conditions without requiring DNS snapshots. ROMs with varying degrees of freedom are validated against DNS in terms of total heat flux, mean profiles, large-scale flow structures, bifurcation diagrams and energy spectra. The objective is to demonstrate that these ROMs can serve as DNS surrogates over a wide range of Rayleigh numbers, setting the stage for future applications in state observation, control and optimization. 
 In our companion paper~\cite{flores2026state}, we demonstrate a practical application of these ROMs for state estimation. The model is combined with an extended Kalman filter (EKF) to track the evolution of the system state across different dynamical regimes from sparse measurements.
 While 3D effects can be important in RB convection, two-dimensional simulations capture the essential dynamics of pattern formation, bifurcations, and heat transport mechanisms~\cite{van2013comparison}. Moreover, 2D configurations allow for extensive parametric studies at a reduced computational cost, making them valuable for ROM development and validation. Although this is not carried out here, an extension of the ROM formulation \CORR{to} a 3D setting with two periodic directions is straightforward.

The remainder of this manuscript is organized as follows. In~\S\ref{sec:theory}, we present the mathematical formulation used to derive the ROMs. The numerical methods employed in this study are described in~\S\ref{sec:numerics}. A comprehensive validation of the ROM against DNS is provided in~\S\ref{sec:results}, including comparisons of mean vertical profiles (\S\ref{sec:profiles}), heat flux scaling with the Rayleigh number (\S\ref{sec:nusselt}), and flow structures (\S\ref{sec:flow_structures}). In~\S\ref{sec:bifurcation_analysis}, we compare dynamical regimes using phase portraits and energy spectra, and exploit the low computational cost of the ROM to perform a detailed characterization of the transition to chaos through Poincaré sections and Lyapunov exponents. Finally, \S\ref{sec:conclusions} summarizes our findings and outlines directions for future research.

\section{Theoretical background}\label{sec:theory}
A two-dimensional flow between two parallel walls is considered. Distances are normalized with the channel height, $h$, so that the wall-normal direction spans from $y=0$ to $y=1$ in dimensionless variables. The fluid domain is periodic in the $x$ direction and extends from $x=0$ to $x=L_x$. In nondimensional form, the mass, momentum and energy conservation equations for a Boussinesq fluid read,
\begin{equation}\label{eq:cont}
\bnabla\bcdot\boldsymbol{u} = 0,
\end{equation}
\begin{equation}\label{eq:mom}
\frac{\partial \boldsymbol{u}}{\partial t} + \boldsymbol{u}\bcdot\bnabla\boldsymbol{u} = -\nabla p + Pr\theta\boldsymbol{e}_y  + \frac{Pr}{\sqrt{Ra}}\bnabla^2\boldsymbol{u} + \boldsymbol{d},
\end{equation}
\begin{equation}
    \label{eq:ener}\frac{\partial \theta}{\partial t} + \boldsymbol{u}\bcdot\nabla\theta = \frac{1}{\sqrt{Ra}}\bnabla^2\theta + q.
\end{equation}
In eqs.~\eqref{eq:cont}--\eqref{eq:ener}, $\boldsymbol{u}$, $\theta$ and $p$ denote the dimensionless velocity, temperature and {driven} pressure, respectively. The terms $\boldsymbol{d}$ and $q$ in eq.~\eqref{eq:mom} and~\eqref{eq:ener}, represent a spatially-distributed time-evolving body force applied on the fluid and a dimensionless heat release rate per unit volume, respectively. {The temperature is non-dimensionalized as $\theta=(T-T_0)/\Delta T$, where $\Delta T= T_1 - T_0$. Here, $T_1$ and $T_0$ denote the imposed dimensional temperatures on the bottom and top wall, respectively. The velocity is normalized by $\kappa\sqrt{Ra}/h$, where $\kappa$ is the thermal diffusivity of the fluid and $Ra$ is the Rayleigh number,
\begin{equation}\label{eq:Ray}
    Ra=  \frac{\sigma\Delta T g h^3}{\nu \kappa}.
\end{equation}
In eq.~\eqref{eq:Ray}, $g$ is the gravitational acceleration, $\nu$ is the kinematic viscosity and $\sigma$ is the thermal expansion coefficient.
The term {$Pr\theta\boldsymbol{e}_y$} represents the upward-pointing buoyancy force responsible for the two-way coupling between the momentum and the energy equation. Here, $Pr=\nu/\kappa$ denotes the Prandtl number. No-slip boundary conditions are applied at $y=0$ and \CORR{$y=L_y=1$} and periodic boundary conditions are considered at $x=0$ and $x=L_x$.}

Equations~\eqref{eq:cont}--\eqref{eq:ener} can be projected onto a set of spatial modes to transform this system of {PDEs} into a system of ODEs that describe the time-evolution of the amplitude coefficients of the modes. In this respect, two approaches can be adopted. On the one hand, it is possible to use two independent modal bases for the velocity and the temperature field. Each of these bases has its own time-dependent scaling coefficients. On the other hand, we can build a {coupled basis of modes where the} velocity and temperature components are scaled by the same amplitude coefficient. 
\subsection{Two-bases approach}
In the uncoupled-bases approach, the velocity field is decomposed as a sum of modes, $\boldsymbol{v}_j$, weighted by an amplitude coefficient, $a_j$,
\begin{equation}
    \boldsymbol{u}(t,\boldsymbol{x}) = \sum_j a_j(t) \boldsymbol{v}_j(\boldsymbol{x}).
\end{equation}
The set of velocity modes, $\boldsymbol{v}_j$, is orthonormal, with an inner product defined as,
\begin{equation}
    \langle\boldsymbol{v}_i,\>\boldsymbol{v}_j\rangle = \frac{1}{L_xL_y}\int_0^{L_x}\int_{0}^{L_y}\boldsymbol{v}_i\bcdot\boldsymbol{v}_j\>dy\>dx= \begin{cases}
        1\quad\mathrm{if}\>i=j\\
        0\quad\mathrm{elsewhere.}
    \end{cases}
\end{equation}
Velocity modes, $\boldsymbol{v}_j$, satisfy boundary conditions and the continuity equation,
\begin{eqnarray}
    \boldsymbol{v}_j(y = 0) &=& 0,\\
    \boldsymbol{v}_j(y = 1) &=& 0,\\
\bnabla\cdot\boldsymbol{v}_j &=& 0.
\end{eqnarray}
The temperature field is decomposed into the sum of a baseline temperature profile, $\theta_0(y)$, and a perturbation, $\theta^\prime(t,\boldsymbol{x})$. The latter is also expressed as sum of modes, $\Theta_j(\boldsymbol{x})$ scaled by a time-evolving amplitude coefficient $b_j(t)$,
\begin{equation}
    \theta(t,\boldsymbol{x}) = \theta_0(y) + \sum_j b_j(t) \Theta_j(\boldsymbol{x}).
\end{equation}
The baseline temperature profile corresponds to the equilibrium conductive solution, $\theta_0(y) = 1 - y$, satisfying $\nabla^2 \theta_0 = 0$. Temperature modes also form an orthonormal {basis} and must satisfy $\Theta_j(y=0) = 0$ and $\Theta_j(y=1) = 0$.
Introducing the Galerkin decomposition into the momentum conservation equation, taking the inner product with $\boldsymbol{v}_i$ and applying the orthonormality property of the modal {basis} gives,
{\begin{equation}\label{eq:Galerkin_momentum}
    \frac{\mathrm{d}a_i}{\mathrm{dt}} + \sum_{j,\>k}\mathcal{N}^u_{ijk} a_k a_j  = Pr\left({\mathcal{F}^u_i} + \sum_j\mathcal{F}^u_{ij} b_j\right) + \frac{Pr}{\sqrt{Ra}}\sum_j\mathcal{D}^u_{ij} a_j + \mathcal{U}^u_i,
\end{equation}}
where,
\begin{eqnarray}
    \label{eq:Gal_mom_NL}\mathcal{N}^u_{ijk} = \langle\boldsymbol{v}_i,\>\boldsymbol{v}_j\cdot\bnabla\boldsymbol{v}_k\rangle,\\
    \label{eq:Gal_mom_For0}{\mathcal{F}^u_i} = \langle\boldsymbol{v}_i, \theta_0\boldsymbol{e}_y\rangle,\\
    \label{eq:Gal_mom_For1}\mathcal{F}^u_{ij} = \langle \boldsymbol{v}_i,\>\Theta_j\boldsymbol{e}_y\rangle,\\
    \label{eq:Gal_mom_Diff}\mathcal{D}^u_{ij} = \langle\boldsymbol{v}_i,\>\bnabla^2\boldsymbol{v}_j\rangle,\\
    \label{eq:Gal_mom_U}{\mathcal{U}^u_i} = \langle\boldsymbol{v}_i,\>\boldsymbol{d}\rangle.
\end{eqnarray}
For divergence-free velocity modes, the pressure term can be eliminated taking into account the boundary conditions. \CORR{In eq.~\eqref{eq:Galerkin_momentum}, $\mathcal{N}^u_{ijk}$ is the nonlinear convective term, {$\mathcal{F}^u_i$} is the forcing owing to the baseline temperature field, $\mathcal{F}^u_{ij}$ is the buoyancy effect generated by temperature perturbations, $\mathcal{D}^u_{ij}$ is the viscous diffusion term and $\mathcal{U}^u_i$ is the projection of the distributed body force onto mode $\boldsymbol{v}_i$}.
Applying the same procedure to the energy equation gives,
\begin{equation}\label{eq:Galerkin_energy}
    {\frac{\mathrm{d}b_i}{\mathrm{dt}} + \sum_j\mathcal{L}^\theta_{ij}a_j  + \sum_{j,\>k}\mathcal{N}^\theta_{ijk} a_k b_j  =  \frac{1}{\sqrt{Ra}}\sum_j\mathcal{D}^\theta_{ij} b_j + \mathcal{U}^\theta_i},
\end{equation}
where,
\begin{eqnarray}
    \label{eq:Gal_ener_NL}\mathcal{N}^\theta_{ijk} = \langle\Theta_i,\>\boldsymbol{v}_k\cdot\bnabla\Theta_j\rangle,\\
    \label{eq:Gal_ener_Lin}\mathcal{L}^\theta_{ij} = \langle\Theta_i,\>\boldsymbol{v}_j\cdot\bnabla\theta_0(y)\rangle,\\
    \label{eq:Gal_ener_Diff}\mathcal{D}^\theta_{ij} = \langle\Theta_i,\>\bnabla^2\Theta_j\rangle,\\
    \label{eq:Gal_ener_U}{\mathcal{U}^\theta_i} = \langle\Theta_i,\>q\rangle.
\end{eqnarray}
\CORR{In eq.~\eqref{eq:Galerkin_energy},  $\mathcal{N}^\theta_{ijk}$ represents the nonlinear advection of temperature perturbations, $\mathcal{L}^\theta_{ij}$ is the linear transport of the baseline conductive temperature profile by the velocity perturbations, $\mathcal{D}^\theta_{ij}$ represents thermal diffusion, and $\mathcal{U}^\theta_i$ is the volumetric heat source term.}
The Galerkin-projection ROM with two independent modal bases is defined by the system of ODEs in eqs.~\eqref{eq:Galerkin_momentum} and~\eqref{eq:Galerkin_energy}. {To complete the formulation}, two independent orthonormal bases for velocity and temperature must be supplied. 
\CORR{A procedure to obtain orthonormal modal bases that satisfy the required boundary conditions is detailed in \S\ref{sec:adjoint}. }

\subsection{Single-basis approach}
A Galerkin-projection ROM is now built using one single orthonormal modal basis where modes include a velocity and a temperature component. First, an extended state vector grouping the velocity and the temperature fields is defined,
\begin{equation}
    \mathcal{X} = \begin{bmatrix}
        u&v&\theta
    \end{bmatrix}^T.
\end{equation}
{The computation of the Gramians via the adjoint equations requires introducing a representative inner product for the coupled modal basis. For a stratified turbulent channel flow,~\citet{ahmed2021resolvent} applied the concept of available potential energy~\cite{lorenz1955available, saltzman1962finite, winters1995available, hughes2013available} to define an inner product for velocity and density fluctuations. Here, we adopt the same approach as~\citet{podvin2001low} and~\citet{soucasse2019proper}, who introduced a scaling for temperature perturbations, $\gamma^2$, to define a coupled inner product. This weight factor represents the ratio between the average energy of the velocity and temperature fields, $E_k$ and $E_\theta$, respectively,}
\begin{equation}\label{eq:orthogonality_coupled}
    {\langle\mathcal{X}_i,\>\mathcal{X}_j\rangle_c = \langle \mathcal{X}_i, W \mathcal{X}_j\rangle =\frac{1}{L_xL_y}\int_0^{L_x}\int_{0}^{L_y}\left(u_iu_j + v_iv_j + \gamma^2\theta_i\theta_j\right)\>dy\>dx,}
\end{equation}
where $W$ is the weight matrix,
\begin{equation}
    {W = \begin{bmatrix} 1&0&0\\0&1&0\\ 0&0 &\gamma^2    \end{bmatrix}.}
\end{equation}
The extended state is decomposed into the sum of the equilibrium conductive solution $\mathcal{X}_0=\begin{bmatrix}0&0&\theta_0(y)\end{bmatrix}^T$ and a time-dependent perturbation, $\mathcal{X}^\prime(t, \boldsymbol{x})$, that is in turn expressed as a sum of orthonormal modes, $\chi_j(\boldsymbol{x})$, satisfying homogeneous boundary conditions,
\begin{equation}\label{eq:one-basis-modal-decomposition}
    \mathcal{X}(t, \boldsymbol{x}) = \begin{bmatrix}
        0&0&\theta_0(y)
    \end{bmatrix}^T + \sum_j c_j(t)\chi_j(\boldsymbol{x}).
\end{equation}
For convenience, the components of the $\chi_j(\boldsymbol{x})$ modes corresponding to the velocity and temperature field are denoted by $\mathcal{V}_j(\boldsymbol{x})$ and $\mathcal{T}_j(\boldsymbol{x})$, respectively. This way, the velocity and temperature fields can be reconstructed as, 
\begin{equation}\label{eq:vel_field_reconstruction_1basis}
    \mathbf{u}(t, \boldsymbol{x}) = \sum_jc_j(t)\mathcal{V}_j(\boldsymbol{x}),
\end{equation}
\begin{equation}\label{eq:temp_field_reconstruction_1basis}
    \theta(t, \boldsymbol{x}) = \theta_0(y) +  \sum_jc_j(t)\mathcal{T}_j(\boldsymbol{x}).
\end{equation}
Note that two arbitrary elements of these subsets are not necessarily orthonormal because the orthonormality condition is only satisfied by the coupled modal basis $\chi_j(\boldsymbol{x})$ under the inner product defined in eq.~\eqref{eq:orthogonality_coupled}.

Introducing the modal decomposition of eq.~\eqref{eq:one-basis-modal-decomposition} into the momentum and energy conservation equations~\eqref{eq:mom}--\eqref{eq:ener} and taking the inner product with $\chi_i$ yields,
{\begin{eqnarray}\label{eq:ROM_1basis}
    \frac{\mathrm{d}c_i}{\mathrm{dt}}  + \sum_j \mathcal{L}^\chi_{ij}c_j +  \sum_{jk}\mathcal{N}_{ijk}^\chi c_k c_j =\\
     \nonumber =Pr\left(\mathcal{F}_i^\chi + \sum_j\mathcal{F}^\chi_{ij}c_j\right) + \frac{Pr}{\sqrt{Ra}}\sum_j\mathcal{D}^\mathcal{V}_{ij}c_j + \frac{1}{\sqrt{Ra}}\sum_j\mathcal{D}^\mathcal{T}_{ij}c_j + \mathcal{U}^\chi_i,
\end{eqnarray}}
where,
\begin{eqnarray}
    \label{eq:Gal_coup_Lin}\mathcal{L}^\chi_{ij} = \langle\chi_i,\>\left(\mathcal{V}_j\cdot\nabla\theta_0(y)\right) \boldsymbol{e}_\theta\rangle_c,\\
    \label{eq:Gal_coup_NL}\mathcal{N}^\chi_{ijk} = \langle\chi_i,\>\mathcal{V}_k\cdot\bnabla\chi_j\rangle_c,\\
    \label{eq:Gal_coup_For0} {\mathcal{F}^\chi_{i}} = \langle\chi_i,\>\theta_0(y)\boldsymbol{e}_v\rangle_c,\\
    \label{eq:Gal_coup_For1} \mathcal{F}^\chi_{ij} = \langle\chi_i,\>\mathcal{T}_j\boldsymbol{e}_v\rangle_c,\\
    \label{eq:Gal_coup_Diff_u}\mathcal{D}^\mathcal{V}_{ij} = \langle\mathcal{V}_i,\>\bnabla^2\mathcal{V}_j\rangle,\\
    \label{eq:Gal_coup_Diff_T}{\mathcal{D}^\mathcal{T}_{ij} = \langle\mathcal{T}_i,\gamma^2 \>   \bnabla^2\mathcal{T}_j\rangle,}\\
    \label{eq:Gal_coup_U}{\mathcal{U}^\chi_i = \langle\chi_i,\>\begin{bmatrix}\boldsymbol{d},& q\end{bmatrix}^T\rangle_c.}
\end{eqnarray}
Here, $\boldsymbol{e}_v = \begin{bmatrix}
    0&1&0
\end{bmatrix}^T$ denotes the unit vector in the wall-normal velocity component of the extended state vector, $\mathcal{X}$, and $\boldsymbol{e}_\theta = \begin{bmatrix}
    0&0&1
\end{bmatrix}^T$ is the unit vector in the temperature component of $\mathcal{X}$. Subscript $-c$ in the inner product operator $\langle\cdot\rangle_c$ is used to indicate the weighted inner product defined in eq.~\eqref{eq:orthogonality_coupled}. The system of ODE in eq.~\eqref{eq:ROM_1basis} defines the Galerkin-projection ROM with a coupled orthonormal basis for temperature and velocity. A procedure to obtain a coupled orthonormal {basis} $\chi_j(\boldsymbol{x})$ is detailed in \S\ref{sec:adjoint}.

\subsection{State-space form and controllability Gramian}\label{sec:adjoint}
It is possible to obtain coupled and uncoupled orthonormal modal bases for velocity and temperature from the controllability Gramian of the linearized Rayleigh-Bénard equations~\cite{bagheri2009input}. This procedure was successfully applied by~\citet{cavalieri2022reduced} to obtain a numerically stable ROM for 3D Couette flow.
First, it is necessary to obtain the state-space form of the linearized equations. For that, the pressure term is eliminated by taking the curl of the momentum equation,
\begin{equation}\label{eq:vor_vec}
    \frac{\partial \boldsymbol{\omega}}{\partial t} + \boldsymbol{u}\cdot\bnabla\boldsymbol{\omega} - \boldsymbol{\omega}\cdot\bnabla\boldsymbol{u} = Pr\bnabla\times\theta\boldsymbol{e}_y + \frac{Pr}{\sqrt{Ra}}\bnabla^2\boldsymbol{\omega} + \bnabla\times\boldsymbol{d}.
\end{equation}
In eq.~\eqref{eq:vor_vec}, $\boldsymbol{\omega}$ denotes the flow vorticity.
In a two-dimensional flow, the only non-zero vorticity component is in the spanwise direction $\boldsymbol{e}_z$, $\boldsymbol{\omega} = \omega\boldsymbol{e}_z$, so that the last term of the left hand side can be eliminated and a scalar equation can be written,
\begin{equation}
    \frac{\partial \omega}{\partial t} + \boldsymbol{u}\cdot\nabla{\omega} = Pr\frac{\partial\theta}{\partial x} + \frac{Pr}{\sqrt{Ra}}\nabla^2{\omega} + \left(\partial_xd_y - \partial_y d_x\right).
\end{equation}
The momentum and energy equations are linearized around a static flow solution, $\boldsymbol{u}_0 = 0$ {($\omega_0=0$)}, with a temperature profile corresponding to the conductive equilibrium solution, $\theta_0(y) = 1 - y$, 
\begin{equation}\label{eq:omega_linearized}
    \frac{\partial \omega^\prime}{\partial t}  = Pr\frac{\partial\theta^\prime}{\partial x} + \frac{Pr}{\sqrt{Ra}}\nabla^2{\omega^\prime} + \left(\partial_xd_y - \partial_y d_x\right),
\end{equation}
\begin{equation}\label{eq:temp_linearized}
    \frac{\partial \theta^\prime}{\partial t} + v^\prime\mathrm{d}_y\theta_0  = \frac{1}{\sqrt{Ra}}\nabla^2{\theta^\prime} + q.
\end{equation}
In eq.~\eqref{eq:omega_linearized} and~\eqref{eq:temp_linearized}, $\omega^\prime$ and $\theta^\prime$ denote the vorticity and temperature perturbations around the baseline solution and $v^\prime$ is the wall-normal velocity.
Introducing a streamfunction for the velocity perturbations, $u^\prime = \partial_y\Psi$, $v^\prime = -\partial_x\Psi$ and taking into account that $\omega^\prime = -\nabla^2\Psi$ the equations can be rewritten as,
\begin{equation}
    \frac{\partial}{\partial t}\left(\nabla^2\Psi\right)  = -Pr\frac{\partial\theta^\prime}{\partial x} + \frac{Pr}{\sqrt{Ra}}\nabla^4{\Psi} - \left(\partial_xd_y - \partial_y d_x\right),
\end{equation}
\begin{equation}
    \frac{\partial \theta^\prime}{\partial t} = \partial_x\Psi\mathrm{d}_y\theta_0  + \frac{1}{\sqrt{Ra}}\nabla^2{\theta^\prime} + q,
\end{equation}
where $\nabla^2 = \partial^2_x + \partial^2_y$ is the Laplacian operator and $\nabla^4 = \partial^4_x + 2\partial^2_x\partial_y^2 + \partial_y^4$ denotes the biharmonic operator. 
{The periodicity of the flow in the $x$ direction allows the solution, $\phi(t,\boldsymbol{x})$, to be decomposed into Fourier modes $\phi(t, \mathbf{x}) = \tilde{\phi}(t,y)e^{ik_x x}$ to give,}
{\begin{equation}\label{eq:linear_Fourier_mom}
    \frac{\partial}{\partial t}\left(\nabla^2\tilde{\Psi}\right)  = -ik_xPr\tilde{\theta} + \frac{Pr}{\sqrt{Ra}}\nabla^4{\tilde{\Psi}} - \left(ik_x\tilde{d_y} - \partial_y \tilde{d_x}\right),
\end{equation}
\begin{equation}\label{eq:linear_Fourier_ener}
    \frac{\partial \tilde{\theta}}{\partial t} = ik_x\mathrm{d}_y\theta_0\tilde{\Psi}  + \frac{1}{\sqrt{Ra}}\nabla^2\tilde{\theta} + \tilde{q}.
\end{equation}}
This system of equations satisfies the following boundary conditions,
\begin{eqnarray}\label{eq:boundary_con}
    \label{eq:bc_dirichlet_Psi}\tilde{\Psi}(y=0) = 0 &\quad& \tilde{\Psi}(y=1) = 0,\\
    \label{eq_bc_neuman_Psi}\partial_y\tilde{\Psi}(y=0)= 0&\quad&\partial_y\tilde{\Psi}(y=1)= 0,\\
    \label{eq_bc_dirichlet_theta}\tilde{\theta}(y=0)=0&\quad&\tilde{\theta}(y=1)=0.
\end{eqnarray}
Arranging the equations into matrix form we obtain,
\begin{equation}
    \frac{\partial}{\partial t}\left(\begin{bmatrix}
        \nabla^2 & 0\\
        0 & 1
    \end{bmatrix}\begin{bmatrix}
        \tilde{\Psi}\\\tilde{\theta}
    \end{bmatrix}\right)=\begin{bmatrix}
        \frac{Pr}{\sqrt{Ra}}\nabla^4& -ik_xPr\\ ik_x\mathrm{d}_y\theta_0 &\frac{1}{\sqrt{Ra}}\nabla^2
    \end{bmatrix}\begin{bmatrix}
        \tilde{\Psi}\\\tilde{\theta}
    \end{bmatrix} + \begin{bmatrix}
        \partial_y &-ik_x & 0\\
        0& 0&1
    \end{bmatrix}\begin{bmatrix}
        \tilde{d_x}\\\tilde{d_y}\\\tilde{q}
    \end{bmatrix}.
\end{equation}
This matrix form enables the identification of the state and input matrices,
\begin{equation}\label{eq:direct_state_matrix}
    A =\begin{bmatrix}
        \frac{Pr}{\sqrt{Ra}}\nabla^{-2}\nabla^{4}& -ik_xPr\nabla^{-2}\\ ik_x\mathrm{d}_y\theta_0 &\frac{1}{\sqrt{Ra}}\nabla^2
    \end{bmatrix},
\end{equation}
\begin{equation}\label{eq:direct_control_matrix}
    B = \begin{bmatrix}
        \nabla^{-2}\partial_y &-ik_x\nabla^{-2} & 0\\
        0& 0&1
    \end{bmatrix},
\end{equation}
with a state vector defined by {$\boldsymbol{z}=\begin{bmatrix}\tilde{\Psi} & \tilde{\theta}\end{bmatrix}^T$} and an input vector, {$\boldsymbol{w} = \begin{bmatrix}\tilde{d}_x & \tilde{d}_y &\tilde{q}\end{bmatrix}^T$}. The observation matrix $C$ depends on the set of observed variables, also referred to as output vector. For an output vector {$\boldsymbol{y}=\begin{bmatrix} \tilde{u}&\tilde{v}&\tilde{\theta} \end{bmatrix}^T$} containing the {Fourier transform of the} two velocity components and the temperature perturbations the observation matrix is given by,
\begin{equation}
    C = \begin{bmatrix}
        \partial_y &0\\
        -ik_x &0\\
        0&1
    \end{bmatrix}.
\end{equation}
To determine the adjoint linear equations, it is necessary to define an inner product in the state space, $\boldsymbol{z}$. It is possible to derive the expression of the inner product in the state space from the inner product definition for the coupled modal basis given in eq.~\eqref{eq:orthogonality_coupled}. {The decomposition into Fourier modes in the $x$ direction allows reducing the inner product to an integral in the $y$ direction only,}
\begin{equation}\label{eq:inner_product_SS}
    \langle\boldsymbol{y}_1,\>W \boldsymbol{y}_2\rangle = \frac{1}{L_y}\int_{0}^{L_y}\boldsymbol{y}_1^HW\boldsymbol{y}_2dy = \int_{0}^{1}\left(C\boldsymbol{z}_1\right)^H W C\boldsymbol{z}_2dy = 
    \int_{0}^{1}\boldsymbol{z}_1^HM\boldsymbol{z}_2dy,
\end{equation}
where superscript $-H$ {represents the conjugate transpose operator}. 
The inner product in the state space is denoted by $\langle\cdot,\cdot\rangle_e$ and defined by,
\begin{equation}\label{eq:inner_prod}
    \langle\boldsymbol{z}_1,\>\boldsymbol{z}_2\rangle_e = \langle\boldsymbol{z}_1,\>M\boldsymbol{z}_2\rangle_{L_2},
\end{equation}
where $\langle\cdot,\>\cdot\rangle_{L_2}$ is the $L^2$ norm in $0\leq y\leq1$ and $M$ is the inner product weight matrix that reads,
\begin{equation}\label{eq:state_inner_prod}
    M = \begin{bmatrix}
        -\nabla^{2}&0\\0&\gamma^2
    \end{bmatrix}.
\end{equation}
In eq.~\eqref{eq:inner_product_SS}, the expression for the $M$ matrix is obtained via integration by parts. The detailed mathematical development is provided in Appendix~\ref{sec:appendix_derivation}.
Using the inner product defined in eq.~\eqref{eq:inner_prod} and taking into account the boundary conditions from eqs.~\eqref{eq:bc_dirichlet_Psi} -- \eqref{eq_bc_dirichlet_theta}, it is possible to compute the adjoint system by solving,
\begin{equation}
    \langle A\boldsymbol{z}_1,\>\boldsymbol{z}_2\rangle_e = \langle \boldsymbol{z}_1,\> A^+\boldsymbol{z}_2\rangle_e,
\end{equation}
\begin{equation}
    \langle \boldsymbol{z},\>B\boldsymbol{w}\rangle_e = \langle B^+\boldsymbol{z},\>\boldsymbol{w}\rangle_{L_2},
\end{equation}
\begin{equation}
    \langle \boldsymbol{y},\>C\boldsymbol{z}\rangle_{L_2} = \langle C^+\boldsymbol{y},\>\boldsymbol{z}\rangle_e.
\end{equation}
Here, $A^+$, $B^+$ and $C^+$ denote the adjoint state, control and observation matrices. This procedure is similar to the one described by~\citet{jovanovic2005componentwise} and~\citet{ilak2008modeling} for a linearized three-dimensional channel flow. The expressions of the adjoint state and control matrices are,
\begin{equation}
    A^+ = \begin{bmatrix}
        \frac{Pr}{\sqrt{Ra}}\nabla^{-2}\nabla^{4}& i\gamma^2 k_x\nabla^{-2}\mathrm{d}_y\theta_0\\ 
        -ik_xPr/\gamma^2 &\frac{1}{\sqrt{Ra}}\nabla^2
    \end{bmatrix},
\end{equation}
\begin{equation}
    B^+ = \begin{bmatrix}
        \partial_y & 0 \\ -ik_x & 0\\ 0&\gamma^2
    \end{bmatrix}.
\end{equation}
The derivation of the adjoint system equations can be found in Appendix~\ref{sec:appendix_derivation}. For this system of equations, the controllability Gramian $\Phi$ can be computed by solving the Sylvester equation,
\begin{equation}\label{eq:sylvester}
    A\Phi + \Phi A^+ + BB^+ = 0.
\end{equation}
The eigenfunctions of the controllability Gramian, $\boldsymbol{z}_\lambda$, provide a basis of {complex-valued} orthogonal modes with an inner product defined by eq.~\eqref{eq:inner_prod},
\begin{equation}
    \Phi \cdot \boldsymbol{z}_\lambda = \lambda\cdot \boldsymbol{z}_\lambda.
\end{equation}
{Each} eigenfunction is transformed into an output vector via multiplication by the observation matrix, $\tilde{\chi}_\lambda=C\boldsymbol{z}_\lambda$. Note that the use of a streamfunction ensures that the velocity modes satisfy the continuity equation. From each complex mode, $\tilde{\chi}_\lambda(y)$, two real modes defined in the $(x,\>y)$ plane are extracted by taking the real and imaginary part of $\tilde{\chi}_\lambda(y)\exp\left(i k_x x\right)$, respectively,
\begin{eqnarray}\label{eq:2D_modes_even}
    \chi_{j} = \Re\left[\tilde{\chi}_\lambda(y)\exp\left(i k_x x\right)\right],\\
    \label{eq:2D_modes_odd}\chi_{j+1} = \Im\left[\tilde{\chi}_\lambda(y)\exp\left(i k_x x\right)\right].
\end{eqnarray}

\CORR{By construction, these two real modes share the same wall-normal structure but are in phase quadrature in the $x$-direction. This phase relationship allows the Galerkin model to represent continuous translations of flow structures along the periodic direction.} 
Finally, the normalization of these modes with their module provides an orthonormal coupled modal basis.
{For $k_x = 0$, corresponding to homogeneous modes in the $x$ direction, the off-diagonal terms in the direct and adjoint state matrices $A$ and $A^+$ are zero and Stokes-diffusion modes are obtained. In this case, two independent modal bases can be built for the velocity and temperature fields using Stokes and thermal diffusion modes, respectively.} 
Indeed, the Stokes-diffusion modes can also be used to obtain nonhomogeneous modes in the $x$-direction. This provides two independent orthonormal modal bases for velocity and temperature for arbitrary values of $k_x$. {We have used this approach to build the uncoupled bases as it corresponds to controllability modes obtained for the uncoupled momentum and energy equations with zero velocity in the base flow (leading to Stokes modes) and the conductive base state (leading to diffusion modes).} Note that in the uncoupled basis approach, the modal basis is independent of the $Ra$ and $Pr$ numbers. In contrast, in the coupled basis approach, the modal basis depends on the generation parameters, i.e. the values of $\gamma^2$, $Ra$ and $Pr$ used to compute the controllability Gramians of the linearized system.

\section{Numerical methods}\label{sec:numerics}
\subsection{ROM generation}
Although Rayleigh-Bénard convection is governed by only two non-dimensional parameters, i.e. the Prandtl and the Rayleigh numbers, the domain aspect ratio affects large-scale flow structures in numerical simulations~\cite{jimenez1991minimal, wang2020zonal}. The choice of the domain aspect ratio, $L_x$, is influenced by the boundary conditions for the velocity imposed at the bottom and top surfaces. 
For free-slip boundary conditions, the critical stability limit is found at a Rayleigh number of $Ra_c=657.5$ for a \CORR{critical wavenumber} of $k_c = \pi/\sqrt{2}$. 
Therefore, in order to capture the linear stability limit, the minimum extension of the domain in the $x$ direction is $L_x=2\sqrt{2}$. A smaller domain cannot fit waves with the critical wavenumber, $k_c$, resulting in a shift of the stability limit towards higher $Ra$ numbers. For no-slip boundaries, the critical stability limit is located at $Ra_c=1707.8$ for a \CORR{wavenumber} of $k_c\simeq 3.117$. Because this value is close to $\pi$, setting the domain length to $L_x=2$ results in a critical Rayleigh number that is similar to the theoretical value predicted by the linear stability analysis of the equations~\cite{getling1998rayleigh}. Since no-slip boundary conditions are applied in this work, the length of the domain in $x$ is set to $L_x=2$.

To build the coupled orthonormal bases, the direct and adjoint system matrices are computed for each wavenumber, $k_x$, after discretization in $y$. 
The $y$ direction is discretized using $n_y=64$ Chebyshev \CORR{modes}. The corresponding differentiation matrices are modified to enforce the boundary conditions from eqs.~\eqref{eq:bc_dirichlet_Psi}--\eqref{eq_bc_dirichlet_theta} as described in~\citet{trefethen2000spectral}.
To compute the controllability Gramians, the Sylvester equation in eq.~\eqref{eq:sylvester} is solved using LAPACK routines~\cite{anderson1999lapack}. In 2D flows, it is not necessary to evaluate negative wavenumbers since modes satisfy $\boldsymbol{v}(k_x) = -\boldsymbol{v}(-k_x)$. Hence, we consider $n_\alpha$ wavenumbers of the form $k_x = j\alpha$ where $j = 0, 1, 2, \ldots, n_\alpha-1$ and $\alpha = 2\pi/L_x$. The number of modes generated for each wavenumber is denoted by $n_\beta$ which results in a total of $N=n_\alpha\times n_\beta$ modes per basis.

In ROMs with coupled modes, the number of degrees of freedom (DoF) of the system is equal to the basis dimension $n=N$ while in uncoupled ROMs it is twice the dimension of the basis $n=2N$. The comparison between coupled and uncoupled ROMs is established using models with the same number of DoF, $n$. 
\CORR{To ensure dimensional equivalence, the coupled models are built with twice the number of modes per wavenumber, $n_\beta$, as the uncoupled models. This adjustment is necessary to resolve a comparable number of physical vertical structures. As noted in \S\ref{sec:adjoint}, representing the continuous translation of a spatial structure requires a pair of real modes in phase quadrature. In the uncoupled ROM, the relative phase between velocity and temperature structures can be adjusted dynamically by their independent amplitude coefficients, $a_i$ and $b_i$, meaning a single complex eigenfunction (two real modes) is sufficient per variable. However, in the coupled approach, velocity and temperature are locked to a single amplitude coefficient, $c_i$. To represent both the spatial translation of the flow and all necessary relative phase shifts between velocity and temperature, the coupled Gramian naturally produces pairs of complex eigenfunctions with similar vertical structures. This generates four real modes per vertical structure, compared to two in the uncoupled case. Therefore, doubling $n_\beta$ in the coupled approach ensures that both models resolve the same underlying physical structures for a given computational cost.}
In this work, coupled bases are generated with $Ra=1$ and $Pr=1$ and the weight factor for \CORR{temperature} perturbations in the inner product, eq.~\eqref{eq:orthogonality_coupled}, is set to $\gamma^2=1.24$. The choice of this value for the scaling factor relies on energy considerations and is explained in detail in \S\ref{sec:gamma2}. Table~\ref{tab:mod_comp} collects the modal structure of the Galerkin models evaluated in this work.

The ROM coefficients are obtained by projecting the equations onto the two-dimensional modes generated after applying eqs.~\eqref{eq:2D_modes_even}--\eqref{eq:2D_modes_odd}. To avoid aliasing in the calculation of the nonlinear terms, the number of grid points in the $x$ direction, which is discretized using a Fourier basis, is set to $4(n_\alpha-1)+2$. This discretization resolves the high-wavenumber components that are generated by convolution of nonlinear terms in modes with the highest spatial frequency component $k_x=(n_\alpha-1)\alpha$.

\begin{figure}
	\nolinenumbers
    \centering
    \includegraphics[width=1\linewidth]{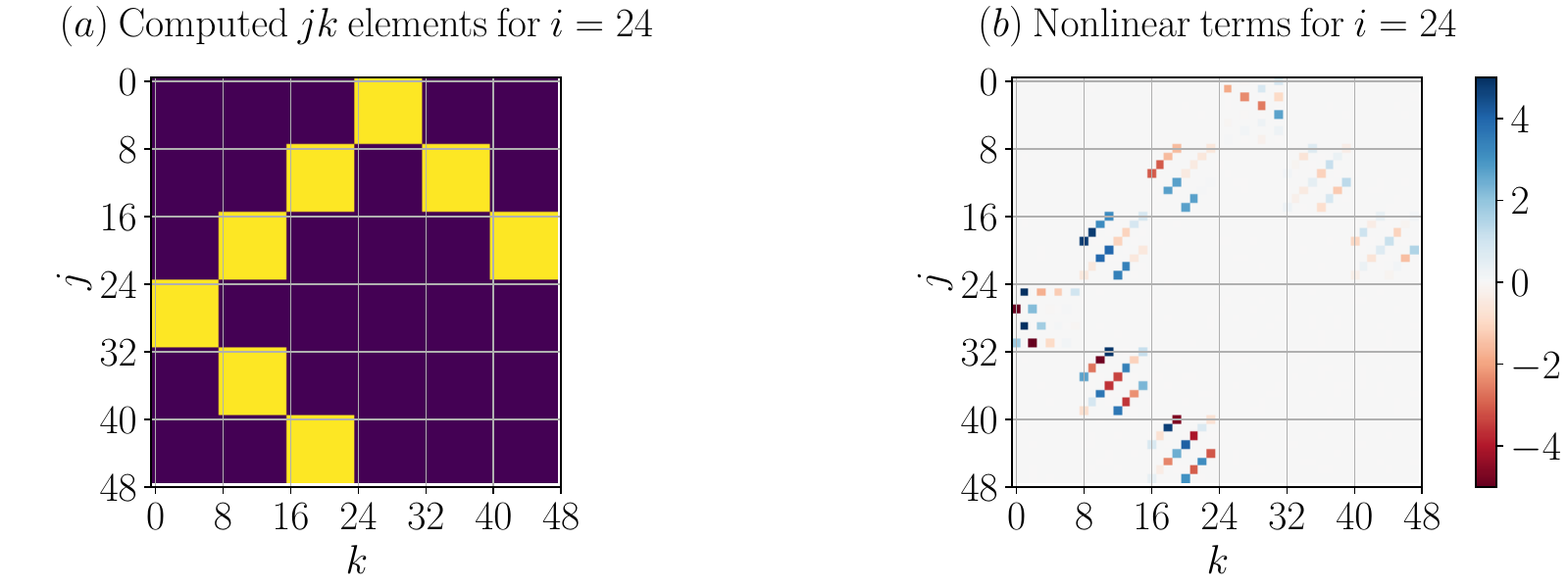}
    \caption{\CORR{Computed elements (a) and sparse structure (b) of the nonlinear term, $\mathcal{N}_{ijk}$.}}
    \label{fig:triad_rule}
\end{figure}

Owing to the nature of triad interactions in incompressible flows~\cite{waleffe1992nature}, only the elements of $\mathcal{N}_{ijk}$ that satisfy $k_i + k_j = k_k$, $k_j + k_k = k_i$ or $k_k + k_i = k_j$ are non-zero.
Here, $k_i$ denotes the wavenumber of the mode with index $i$. As a result, only a fraction of the elements of $\mathcal{N}_{ijk}$ needs to be computed. This is illustrated in Fig.~\ref{fig:triad_rule} for {the uncoupled model U96 with $n_\alpha=6$ and $n_\beta=8$}. In {Fig.~\ref{fig:triad_rule}a}, the $jk$-elements of $\mathcal{N}_{i=24}$ satisfying the triad rule are represented in yellow. {Figure~\ref{fig:triad_rule}b} displays the value of the elements of $\mathcal{N}_{i=24}$. 
{This substantially reduces the number of operations to be performed in the computation of the nonlinear term. The computational savings depend on the wavenumbers considered, but they generally increase with \CORR{the} number of wavenumbers in the ROM. For instance, the number of projecting operations in the nonlinear term is reduced by $78.7\%$ in model U96 ($n_\alpha=6$), by $83.4\%$ in U192 ($n_\alpha=8$) and by $88.5\%$ in U384 ($n_\alpha=12$).}
This property also allows us to treat the $\mathcal{N}_{ijk}$ tensor as a sparse matrix with a $n_\beta\times n_\beta$ block structure. This reduces the memory usage and speeds up the calculations in ROMs with a relatively large number of modes.

In the Galerkin ROMs presented in this work, mass conservation is ensured by using a streamfunction formulation to generate individual modes, which inherently satisfies the continuity equation. Boundary conditions are enforced through appropriate manipulation of the discretized differential operators, ensuring that each mode satisfies the prescribed boundary constraints. The use of DNS-level resolution in the $y$ direction, combined with spectral differentiation operators and carefully selected resolution in $x$ to prevent aliasing, enables us to achieve lossless quadratic terms with $\sum_{ijk}\mathcal{N}_{ijk} \sim \mathcal{O}(10^{-13})$, thereby preserving energy conservation. 
The discussion of the numerical properties of these Galerkin ROMs is completed by the results presented in \S\ref{sec:results}, where ROMs with increasing DoFs are compared to DNS at different $Ra$ numbers.

Finally, the polynomial form of Galerkin ROMs allows for an analytical expression of the system Jacobian. This has been exploited in previous works to compute fixed-point solutions and orbits~\cite{mccormack2024multi}. Here, this feature is used to reduce the computational time required to integrate the ROM equations through implicit integration methods. In the uncoupled ROM, the Jacobian of the system reads,
\begin{equation}\label{eq:gradient}
   \CORR{ \begin{bmatrix}\dfrac{\partial \dot{a}_i}{\partial a_j}   & \dfrac{\partial \dot{a}_i}{\partial b_j} \\ 
    \dfrac{\partial \dot{b}_i}{\partial a_j}                  & \dfrac{\partial \dot{b}_i}{\partial b_j} \end{bmatrix}} = \begin{bmatrix} \frac{Pr}{\sqrt{Ra}}\mathcal{D}_{ij}^u  -  \sum_{k}\mathcal{N}_{ijk}^u a_k -  \sum_{j}\mathcal{N}_{ijk}^u a_j & Pr \mathcal{F}_{ij}^u\\
    - \mathcal{L}^\theta_{ij} -\sum_{j}\mathcal{N}_{ijk}^\theta b_j & \mathcal{D}^\theta_{ij}/\sqrt{Ra}  -  \sum_{k}\mathcal{N}_{ijk}^\theta a_k    \end{bmatrix}.
\end{equation}
In the coupled ROM approach, the Jacobian expression is,
\begin{equation}\label{eq:gradient_U}
    \CORR{\frac{\partial \dot{c}_i}{\partial c_j}} = - \mathcal{L}^\chi_{ij} -  \sum_{k}\mathcal{N}_{ijk}^\chi c_k -  \sum_{k}\mathcal{N}_{ijk}^\chi c_j  +  Pr\mathcal{F}^\chi_{ij} + (Pr/\sqrt{Ra})\mathcal{D}^\mathcal{V}_{ij} + (1/\sqrt{Ra})\mathcal{D}^\mathcal{T}_{ij}.
\end{equation}
In this work, an Adams/BDF method with automatic stiffness detection~\cite{hindmarsh1983odepack, petzold1983automatic} is used to integrate the ROM equations. The use of this scheme reduces the computation time with respect to an explicit third-order Runge-Kutta method across a wide range of simulation parameters. 

\begin{table}
\centering
\nolinenumbers
{\begin{tabular}{ccccccccccc}
Type                       & Nomenclature & $n_{\alpha}$ & $n_{\beta}$ & $N$ & $n$ & $n_x$ & $n_y$               & $Ra$               & \multicolumn{1}{l}{$Pr$}  & $\gamma^2$ \\ \hline
\multirow{3}{*}{Uncoupled} & U96           & 6            & 8          & 48  & 96  & 22    & 64       & \multirow{3}{*}{N/A} & \multirow{3}{*}{N/A}        & \multirow{3}{*}{N/A}\\
                           & U192          & 8            & 12         & 96  & 192 & 30    & 64       &                      &                            & \\
                           & U384          & 12           & 16         & 192 & 384 & 46    & 64       &                      &                            & \\\hline
\multirow{3}{*}{Coupled}   & C96           & 6            & 16         & 96  & 96  & 22    & 64       & \multirow{3}{*}{1}   & \multirow{3}{*}{1}           & \multirow{3}{*}{1.24}\\
                           & C192          & 8            & 24         & 192 & 192 & 30    & 64       &                      &                             & \\
                           & C384          & 12           & 32         & 384 & 384 & 46    & 64       &                      &                             &  \\\hline
\end{tabular}
 \caption{Modal composition, generation parameters and mesh resolution of the different models evaluated. N/A stands for Not Applicable.}
  \label{tab:mod_comp}}
\end{table}

\subsection{Direct Numerical Simulations}

ROM simulations are compared to Direct Numerical Simulations (DNS) for model validation. DNS are performed with the aid of the Python library \textsc{Dedalus}~\cite{burns2020dedalus} for spectral numerical methods. In DNS, $n_x=128$ grid points are used in the $x$ direction which is discretized using a Fourier basis. In the $y$ direction, $n_y=64$ Chebyshev \CORR{modes} are taken. The dealias factor in both bases is $3/2$ and the time integration is performed using an \CORR{implicit-explicit Runge-Kutta method of order 3~\cite{ascher1997implicit}}. If not stated otherwise, simulations are initialized from a randomly perturbed conductive solution with $\theta_0=1-y$. {The input control terms $\mathbf{d}$ and $q$ are set to zero in DNS.}
Equations~\eqref{eq:cont}--\eqref{eq:ener} are integrated over a simulation time of {$T = 500Pr^{-1/2}$}. After neglecting initial transients, all results are reported for statistically steady heat flux conditions.
Table~\ref{tab:DNS} summarizes the DNS results for different Rayleigh numbers and the two Prandtl numbers investigated in this work, $Pr=1$ and $Pr=10$. The mean Nusselt number is defined in eq.~\eqref{eq:Nusselt}. The Reynolds number, {$Re_\tau= \sqrt{Ra}\,\tilde{u}_\tau/Pr$}, is based on the dimensionless friction velocity, $\tilde{u}_\tau$, which is calculated from the maximum viscous stress at the wall, \CORR{$\tilde{u}_\tau=(Pr\max(|\partial_y u|_{w})/\sqrt{Ra})^{1/2}$}. Finally, {$y^+_w = y_w \tilde{u}_\tau\sqrt{Ra}/Pr$} is the dimensionless distance to wall of the second Chebyshev collocation point.

\begin{table}
	\centering
	\nolinenumbers
	\begin{tabular}{cc|ccc|ccc}
		& & \multicolumn{3}{c|}{$Pr=1$} & \multicolumn{3}{c}{$Pr=10$} \\
		$Ra$ & $\R$ & $\langle Nu \rangle$ & $Re_{\tau}$ & $y_w^+$ & $\langle Nu \rangle$ & $Re_{\tau}$ & $y_w^+$ \\ \hline
		2\>400      &  1.4    & 1.43                 & 9.1   & 0.012  				& 1.43 & 2.9 &  0.004 \\
		4\>000      &  2.3    & 1.93                 & 12.8  & 0.017  				& 1.92 & 3.9 &  0.005\\
		8\>000      &  4.7    & 2.48                 & 17.7  & 0.024  				& 2.45 & 5.2 &  0.007 \\
		16\>000     &  9.4    & 2.77                 & 18.4  & 0.025  & 2.80 & 5.9 &  0.008 \\
		{17\>078}     & {10.0}    & {2.83}                 & {19.0}  & {0.026}  & 2.87 & 6.1 & 0.008 \\
		40\>000     & 23.4    & 3.76                 & 27.5  & 0.037  & 3.72 & 8.4 &  0.011 \\
		80\>000     & 46.8    & 4.71                 & 50.2  & 0.068  & 4.30 & 12.2 & 0.016\\
		{170\>777}    & {100.0}    & {5.74}                 & {69.8}  & {0.095}  & 4.95 & 16.3 & 0.022 \\
		400\>000    & 234.2   & 7.03                 & 99.1  & 0.134  & 5.96 & 22.0 & 0.030 \\
		800\>000    & 468.4   & 7.40                 & 118.4 & 0.160  & 7.17 & 28.5 & 0.039 \\
		{853\>883}    & {500.0}   & {7.46}                 & {128.0} & {0.173}  & 7.32 & 29.5 & 0.040 \\
		8\>000\>000 & 4684.5  & 13.22                & 308.4 & 0.418  & 14.03 & 74.3 & 0.101
	\end{tabular}
	\caption{Reduced Rayleigh number, $\R=Ra/Ra_c$, mean vertical heat flux, $\langle Nu\rangle$, Reynolds number, $Re_\tau$, and normalized distance to the wall of the second node, $y^+_w$, in DNS for different $Ra$ numbers and two Prandtl numbers, $Pr=1$ and $Pr=10$.}
	\label{tab:DNS}
\end{table}

\subsection{Inner product scaling factor}\label{sec:gamma2}
\begin{figure}
    \centering
    \nolinenumbers
    \includegraphics[width=0.5\linewidth]{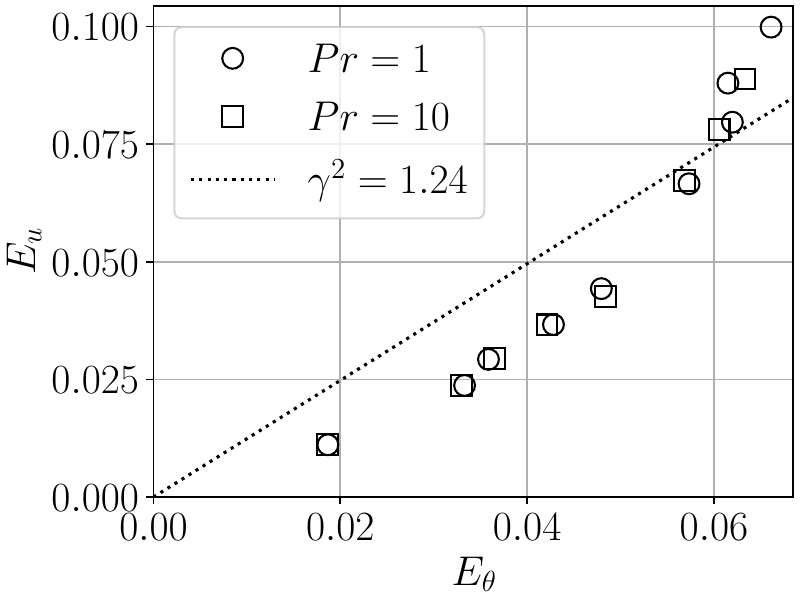}
    \caption{Energy of velocity fluctuations versus temperature perturbations energy for $2.4\times10^3<Ra<8\times 10^6$ and two Prandtl numbers, $Pr=1$ and $Pr=10$. }
    \label{fig:EkEt}
\end{figure}

Before analyzing ROM performance, we justify the choice of the inner product scaling factor, $\gamma^2$. To adjust the $\gamma^2$ factor we follow a similar procedure to that of~\citet{soucasse2019proper}. The total energy of velocity and temperature perturbations, $E_k$ and $E_\theta$, is defined as,
\begin{equation}
	E_k = \left\langle\frac{1}{L_x L_y}\int_0^{L_x}\int_0^{L_y} \boldsymbol{u}^{\prime 2} \, \mathrm{d}y\,\mathrm{d}x\right\rangle, \qquad E_\theta = \left\langle\frac{1}{L_x L_y}\int_0^{L_x}\int_0^{L_y} \theta^{\prime 2} \, \mathrm{d}y\,\mathrm{d}x\right\rangle,
\end{equation}
where $\langle\cdot\rangle$ denotes time-averaging. The scaling factor is then given by $\gamma^2={E_k}/{E_\theta}$. Perturbation energies have been computed from the DNS cases listed in Tab.~\ref{tab:DNS}. Figure~\ref{fig:EkEt} shows $E_k$ versus $E_\theta$ for $2.4\times10^3<Ra<8\times 10^6$ at two Prandtl numbers, $Pr=1$ and $Pr=10$. A least squares fit (black dotted lines) yields $\gamma^2 = 1.24$, which is used throughout this study. This value is in reasonable agreement with $\gamma^2 = 1.303$ reported by~\citet{soucasse2019proper}, and we verified that using their value produces similar results.

\section{Results}\label{sec:results}
Coupled and uncoupled ROMs with different DoFs are now compared to DNS results in terms of mean profiles, total heat flux, flow structures, spectral characteristics, and system dynamics. 
The purpose is to systematically characterize ROM performance as a function of the Rayleigh number and model dimension, $n$, \CORR{to identify} the parameter regimes where these models can serve as reliable surrogates for DNS.
\CORR{Using the DNS dataset with $Pr=1$ (cf. Tab.~\ref{tab:DNS}), we first evaluate the mean vertical profiles in \S\ref{sec:profiles}, followed by the analysis of the total heat flux scaling in \S\ref{sec:nusselt}. The ability of the models to capture large-scale flow structures is assessed in \S\ref{sec:flow_structures}. Finally, \S\ref{sec:bifurcation_analysis} presents a detailed bifurcation analysis, including phase portraits and power spectra, to verify that the ROMs capture the correct dynamical regimes.}

\subsection{Mean profiles}\label{sec:profiles}

\begin{figure}
    \centering
    \nolinenumbers
    \includegraphics[width=0.90\linewidth]{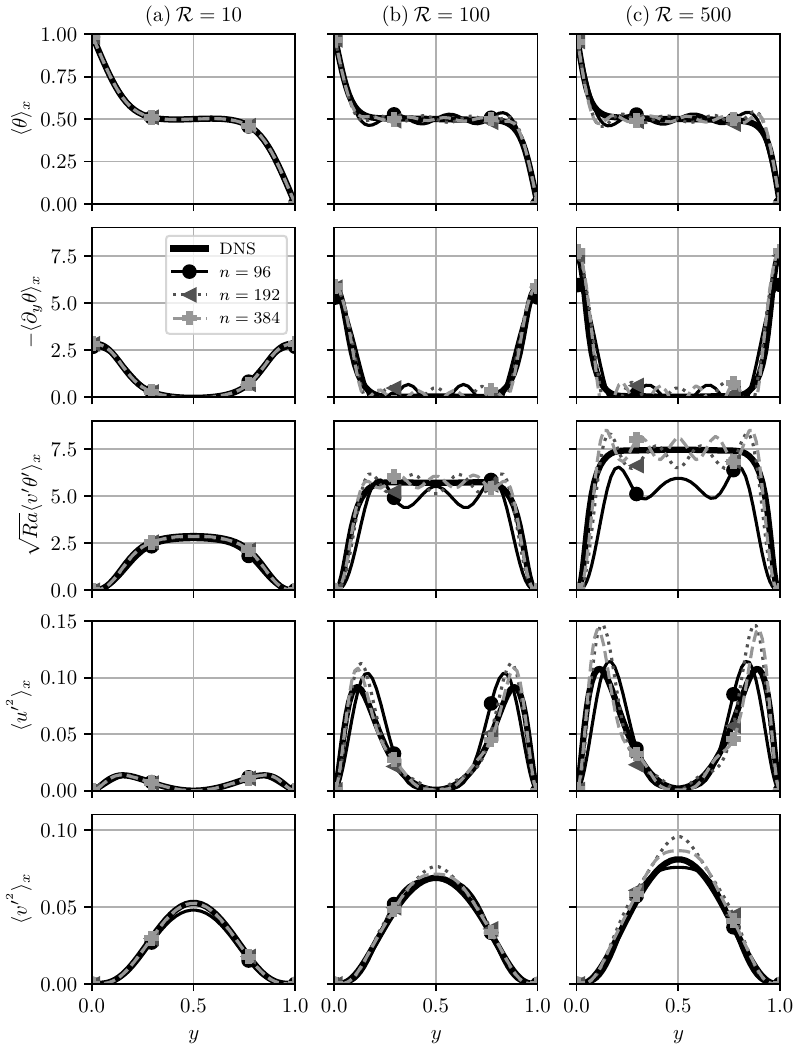}
    \caption{\CORR{Mean vertical profiles of (a) temperature, (b) conductive and (c) convective heat flux, and (d)--(e) normal Reynolds stresses in DNS and models C96 -- C384 for $\R=10$, $\R=100$ and $\R=500$ at $Pr=1$.}}
    \label{fig:profiles_coupled}
\end{figure}

\begin{figure}
    \centering
    \nolinenumbers
    \includegraphics[width=0.9\linewidth]{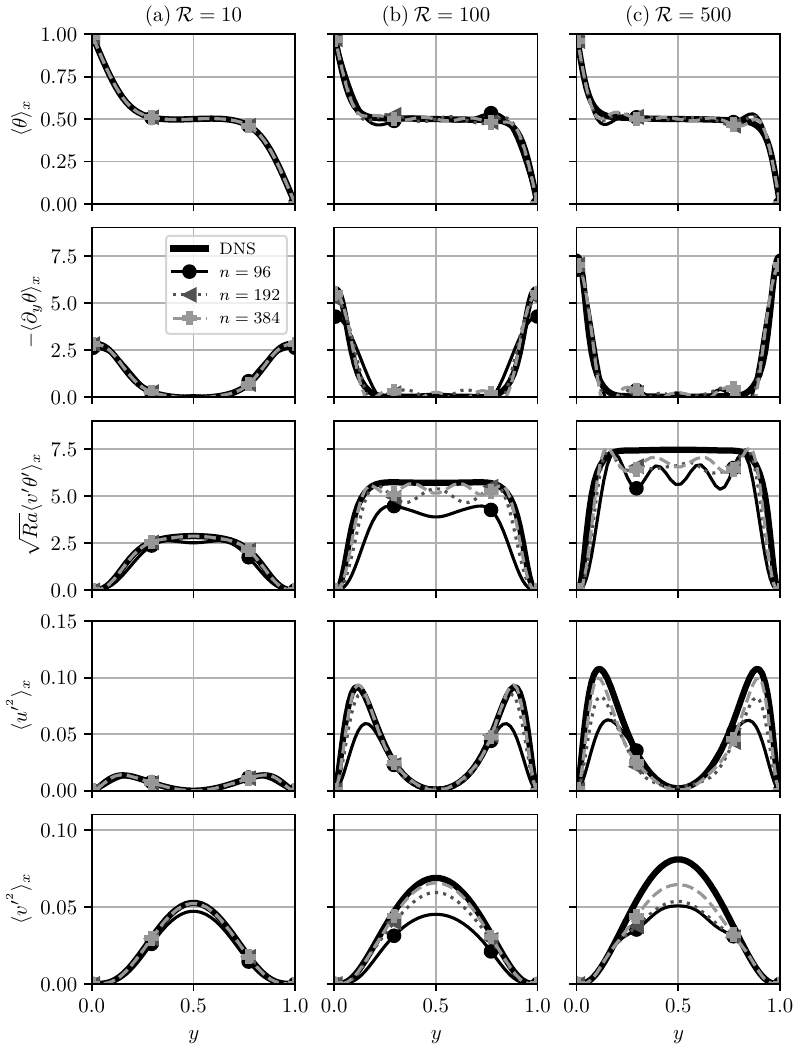}
    \caption{\CORR{Mean vertical profiles of (a) temperature, (b) conductive and (c) convective heat flux, and (d)--(e) normal Reynolds stresses in DNS and models U96 -- U384 for $\R=10$, $\R=100$ and $\R=500$ at $Pr=1$.}}
    \label{fig:profiles_uncoupled}
\end{figure}

ROM results are first compared to DNS in terms of mean vertical profiles of temperature, $\langle\theta\rangle_x$, conductive heat flux, $-\langle\partial_y\theta\rangle_x$, convective heat flux, $\sqrt{Ra}\langle v^\prime\theta^\prime\rangle_x$,  and normal Reynolds stresses, $\langle u^{\prime^2}\rangle_x$ and $\langle v^{\prime^2}\rangle_x$, for various reduced Rayleigh numbers, $\R=Ra/Ra_c$, at $Pr=1$.
ROMs are integrated until the system reaches a statistically steady state. To obtain the mean vertical profiles, the fluid variables are averaged first in the $x$-direction and then in time. \CORR{The agreement between ROM and DNS results is quantified using the relative $L_2$ norm of the error between the mean profiles and defined as $\varepsilon_\phi=||\phi_\mathrm{ROM}-\phi_\mathrm{DNS}||_2/||\phi_\mathrm{DNS}||_2$ for a generic mean profile $\phi(y)$. }

\CORR{In Fig.~\ref{fig:profiles_coupled}, the mean vertical profiles obtained with models C96--C384 are compared to the DNS results at $\R=10$, 100 and 500. The mean temperature profiles, $\langle\theta\rangle_x$ (shown in Fig.~\ref{fig:profiles_coupled}a) are close to DNS in all cases and the differences across the three coupled ROMs are not very remarkable. The errors are below 2\% at $\R=10$ and between 5.7\% and 6.4\% for $\R=500$. The conductive heat flux profiles, $-\langle\partial_y\theta\rangle_x$, are displayed in Fig.~\ref{fig:profiles_coupled}b. At $\R=10$, all the three ROMs are in excellent agreement with the DNS results. The agreement decreases with the Rayleigh number. At $\R=500$, model C96 fails to predict the conductive heat flux at the wall while the other two maintain a reasonable agreement with the DNS. The convective heat flux profiles, $\sqrt{Ra}\langle v'\theta'\rangle_x$, are shown in Fig.~\ref{fig:profiles_coupled}c. At $\R=10$ the $\sqrt{Ra}\langle v'\theta'\rangle_x$ profile is well reproduced by the three coupled ROMs with a 10\% error in the $n=96$ model and about 1\% in models C192 and C384. At $\R=100$, model C96 underpredicts the mid-domain region and the solution oscillates around the DNS profile. At $\R=500$ the oscillations around the smooth DNS profile become more pronounced and the relative errors rise to 32\%, 13\% and 11\% for models C96, C192 and C384, respectively. The coupled ROMs C192 and C384  capture the overall shape of the profile, but C96 does not reproduce the central zone. Figure~\ref{fig:profiles_coupled}d shows the streamwise Reynolds stress profiles, $\langle u'^2\rangle_x$ , at different $\R$ numbers. The three coupled ROMs are in good agreement with the DNS profile at $\R=10$. At $\R=100$, all three models slightly overpredict the peak amplitude. In model C96 the peak location is also shifted towards the center of the domain. At $\R=500$ the peak amplitude is still overpredicted and the errors rise to 32\%, 28\% and 22\%. In Fig.~\ref{fig:profiles_coupled}e, the wall-normal Reynolds stress profiles, $\langle v'^2\rangle_x$, are shown for different $\R$ numbers. At low Rayleigh numbers, there is a good agreement between the ROMs and the DNS, with errors below 10\% for $\R=10$ and $\R=100$. At $\R=500$, the maximum error is about 15\% but the overall shape of the profile is well captured by the three coupled ROMs.
	
The results for uncoupled ROMs U96 -- U384 are shown in Fig.~\ref{fig:profiles_uncoupled}.  For the temperature profiles (Fig.~\ref{fig:profiles_uncoupled}a) DNS and ROM results agree closely with only minor oscillations at high $\R$ numbers. The error is smaller than $5\%$ in all cases. Regarding the conductive heat flux (Fig.~\ref{fig:profiles_uncoupled}b), there is a very good agreement at $\R=10$. When the Rayleigh number is increased to $\R=100$, model U96 tends to underpredict the heat flux at the wall but the three uncoupled ROMs fairly capture the overall shape of the mean profile. At high $\R$ number, the three ROMs maintain a good agreement with the DNS and the profiles are less wavy than in the coupled ROMs. The convective heat flux profiles are displayed in Fig.~\ref{fig:profiles_uncoupled}c. At $\R=10$ all the three uncoupled ROMs agree closely with the DNS results. The error is about $12\%$ in model U96 and below $2\%$ for U192 and U384. At higher $\R$ numbers, the uncoupled models fail to capture the flat mid-domain region and tend to underpredict the convective heat flux. The streamwise Reynolds stress, $\langle u'^2\rangle_x$ is well predicted by the three uncoupled ROMs for $\R=10$.  At $\R=100$, models U192 and U384 show a very good agreement with the DNS results while model U96 underpredicts the profile peaks. When the $\R$ number increases to 500, only model U384 is able to capture the peak amplitude and its location. Finally, Fig.~\ref{fig:profiles_uncoupled}e shows the wall-normal Reynolds stress profiles at different $\R$ numbers. Overall, there is a good agreement in the mean profiles of $\langle v'^2\rangle_x$ but the uncoupled ROMs fail to predict the profile peak at high $\R$ numbers.}

The results presented in Fig.~\ref{fig:profiles_coupled} and Fig.~\ref{fig:profiles_uncoupled} indicate that the mean vertical temperature profile, $\langle\theta\rangle_x$, is captured with great accuracy by large truncations of the system even at high $\R$ numbers. \CORR{For all the ROMs analyzed, the error in mean temperature profile is always below $7\%$.}
In contrast, the largest errors occur for the convective heat flux, $\sqrt{Ra}\langle v^\prime\theta^\prime\rangle_x$, where the ROM profiles tend to be wavy.

Figures~\ref{fig:profiles_coupled} and \ref{fig:profiles_uncoupled} also allow for a direct comparison between the coupled and uncoupled ROM approaches. At low $\R$ numbers both approaches provide similar results. As the Rayleigh number increases, coupled ROMs exhibit better agreement with DNS profiles, particularly for the convective heat flux and the normal Reynolds stresses. While coupled models show more pronounced oscillations around the DNS solution, they capture the overall profile shapes more accurately than uncoupled ROMs. \CORR{In terms of error metrics, the most noticeable difference in accuracy between the coupled and the uncoupled approach is obtained for the $n=96$ ROMs at high $\R$ numbers: $19.2\%$ error (C96) versus $34.7\%$ (U96) for the conductive heat flux and $6.9\%$ error (C96) versus $34.3\%$ (U96) for the wall normal Reynolds stress  at $\R=100$, }

Both coupled and uncoupled ROMs quantitatively reproduce the mean vertical profiles of temperature and heat flux at low Rayleigh numbers. As expected, as the $\R$ number increases, a larger number of modes is required for \CORR{quantitative} agreement, but the main trends remain reasonably captured even with substantial system truncations. Coupled ROMs are found to maintain better agreement with DNS as the Rayleigh number increases. The agreement between DNS and ROMs observed here is consistent with the results of~\citet{cavalieri2022reduced} for three-dimensional Couette flow.

\subsection{Total heat flux}\label{sec:nusselt}
\begin{figure}
     \centering
     \nolinenumbers
     \begin{subfigure}{0.48\textwidth}
         \centering
         \caption{Coupled ROMs}
         \includegraphics[width=1\textwidth]{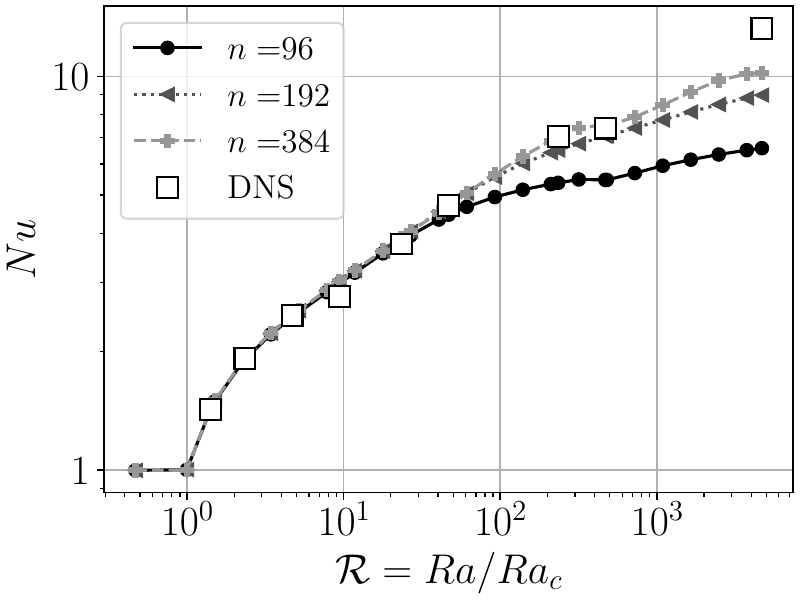}
         \label{fig:Nu_Ra_coupled}
     \end{subfigure}
     \hfill
     \begin{subfigure}{0.48\textwidth}
         \centering
         \caption{Uncoupled ROMs}
         \includegraphics[width=1\textwidth]{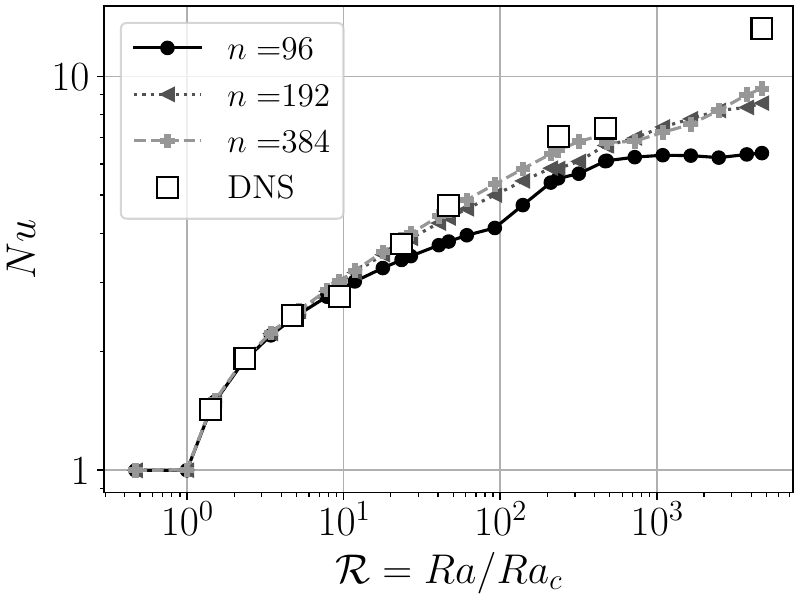}
         \label{fig:Nu_Ra_uncoupled}
     \end{subfigure}     
    \caption{Variation of the $Nu$ number with the $\R=Ra/Ra_c$ number for $Pr=1$. {Axes are in logarithmic scale.}}
    \label{fig:Nusselt_Ra}
\end{figure}

We now assess the ability of ROMs to predict integral quantities of the flow. In RB convection, the integral quantity of interest is the mean vertical heat flux across the fluid layer~\cite{moore1973two}.  The heat flux is characterized using the Nusselt number which is defined as,
\begin{equation}\label{eq:Nusselt}
    {Nu = \sqrt{Ra}\langle v^\prime \theta^\prime\rangle_x - \langle\partial_y \theta\rangle_x},
\end{equation}
and it is \CORR{equal to unity} in the absence of convection. To compute the variation of the heat flux with the Rayleigh number, ROMs are integrated until the $Nu$ reaches a statistically steady value. The time-averaged Nusselt number is then calculated, yielding a single value for each $Ra$, \CORR{since} $Nu$ is uniform in the wall-normal direction. The evolution of $Nu$ as a function of $\R$ is plotted in Fig.~\ref{fig:Nusselt_Ra} for the ROMs in Tab.~\ref{tab:mod_comp} at $Pr=1$. For the sake of comparison, DNS results are represented by white squares.

The results obtained with the coupled ROMs C96--C384 are plotted in Fig.~\ref{fig:Nu_Ra_coupled}. The three coupled models are capable of predicting the critical transition from conduction to convection at $\R=1$. For the three coupled ROMs, a very good agreement with DNS results is obtained up to $\R=23.4$ ($Ra=40\>000$).  Then, the performance of ROMs improves with the number of modes, $n$. In model C384, the error in the Nusselt number with respect to DNS results is $0.23\%$ at $\R=234.2$ ($Ra=400\>000$) and rises to $1.23\%$ at $\R=468.4$ ($Ra=800\>000$). These results show that Galerkin ROMs with significant truncation levels can yield quantitative results for $Ra= 10^5$--$10^6$.

In Fig.~\ref{fig:Nu_Ra_uncoupled}, the evolution of the Nusselt number with the Rayleigh number is represented for the uncoupled ROMs U96–U384, together with the DNS results. As with coupled systems, all three uncoupled-basis ROMs capture the Rayleigh-Bénard instability at $Ra_c$ and maintain excellent agreement with DNS up to $\R=4.7$ ($Ra=8\,000$), where deviations become noticeable for model U96. The comparison between the two projection approaches reveals that coupled models show better performance than uncoupled ROMs giving more accurate results for the same truncation level. This suggests that coupled bases better capture the interactions between velocity and temperature in buoyancy-driven convection, with a more efficient representation of coherent structures in this flow.

Although the accuracy of ROMs degrades with the $Ra$ number, both coupled and uncoupled ROMs maintain numerical stability up to $\mathcal{R}=Ra/Ra_c=10^6$ for integration times of $T=10^4$. We did not evaluate stability beyond this Rayleigh number, since it falls well outside the range where any of the presented models yields accurate results. As in~\citet{cavalieri2022reduced}, this numerical robustness is achieved without any explicit closure model, in contrast to previous POD-Galerkin studies for buoyancy-driven convection in cavities that required additional dissipation terms to prevent numerical blow-up \citep{podvin2001low,podvin2012proper,soucasse2019proper}. 

It is important to note that the choice of normalization is key for the stability of coupled ROMs. Following~\citet{soucasse2019proper}, eqs.~\eqref{eq:cont}--\eqref{eq:ener} were nondimensionalized using the characteristic time $t_c = h^2 / (\kappa \sqrt{Ra})$. With this scaling, the magnitude of dimensionless velocities remains nearly constant with $Ra$ and of the same order as the temperature perturbations, $u \sim \theta^\prime$.  In other works (\citet{howle1996comparison, zienicke1998bifurcations, winchester2022onset}), the thermal diffusion time, $t_c = h^2 / \kappa$, is used instead. That scaling leads to $u \sim \sqrt{Ra}$, which, in coupled ROMs where the same coefficient, $c_i$, applies to both velocity and temperature fields, was seen to result in numerical instabilities and nonphysical behavior at high $Ra$.

Our DNS show that in 2D RB convection with no-slip walls the Nusselt number varies approximately as $Nu\propto\R^{1/4}$ for $1<\R<10^3$. 
These results are in close agreement with those obtained by~\citet{johnston2009comparison} in the same numerical configuration as the one used here. These authors compared the scaling of the Nusselt with the Rayleigh number in 2D simulations with $Pr=1$ and no-slip walls using different thermal boundary conditions. They found that the Nusselt number scales approximately as $\R^{0.284}$ for $10^{6}<Ra<10^{10}$ in domains with fixed-temperature no-slip walls and aspect ratio $L_x=2$. Both the uncoupled and coupled ROMs capture the scaling of the Nusselt number with the reduced Rayleigh number $\R$ over {two} orders of magnitude. This indicates that they could be reliably used to predict the total heat flux across the fluid layer in heat transfer applications involving RB convection.

\subsection{Large-scale flow structures}\label{sec:flow_structures}

\begin{figure}
    \centering
    \nolinenumbers
    \includegraphics[width=1\linewidth]{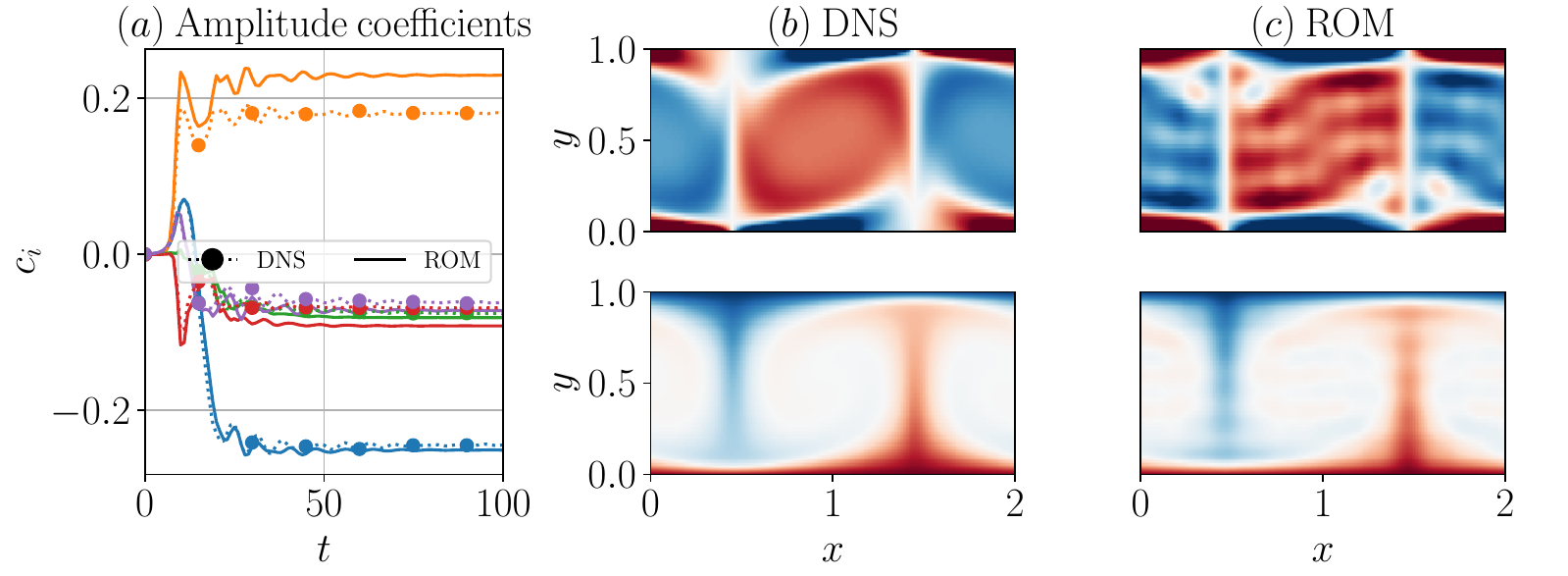}
    \caption{(a) Time-evolution of amplitude coefficients, $c_i(t)$, in DNS and ROM (C384) for $\R=100$ and $Pr=1$. Vorticity and temperature fields at equilibrium in DNS (b) and ROM (c).}
    \label{fig:LSFS_r100_C384}
\end{figure}

\begin{figure}
	\centering
	\nolinenumbers
	\includegraphics[width=1\linewidth]{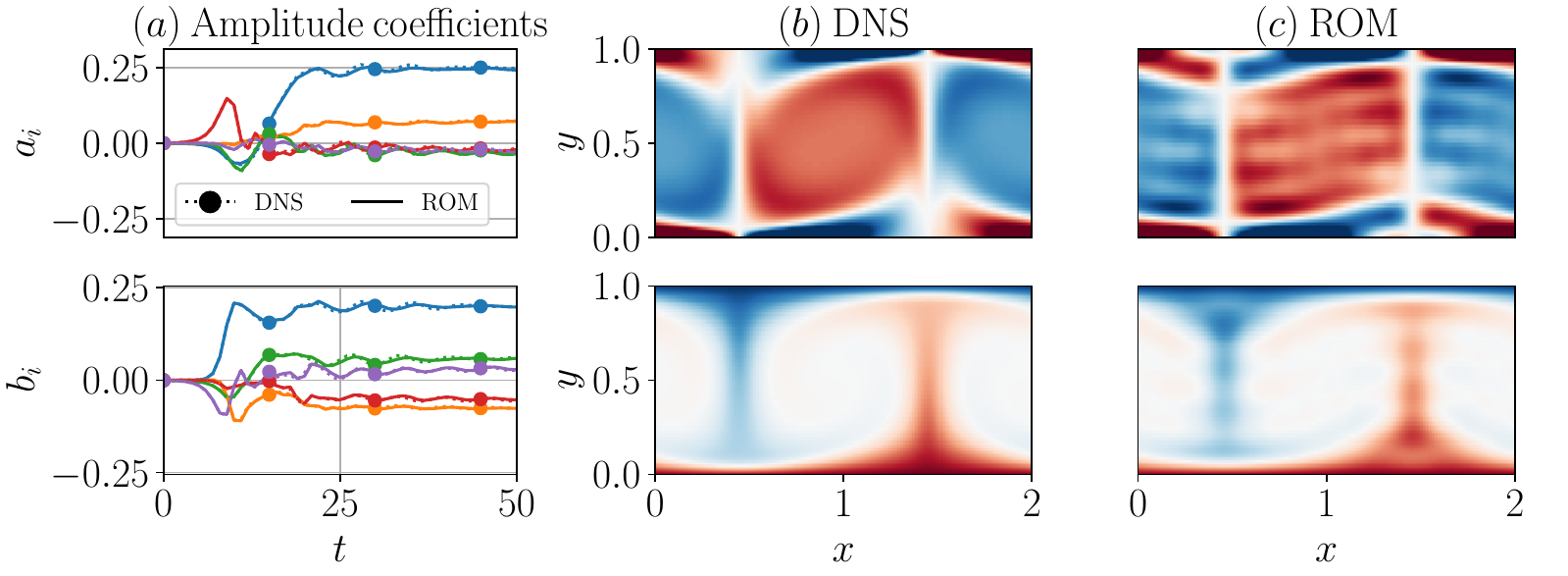}
	\caption{(a) Time-evolution of velocity, $a_i(t)$, and temperature, $b_i(t)$, coefficients in DNS and ROM (U384) for $\R=100$ and $Pr=1$. Vorticity and temperature fields at equilibrium in DNS (b) and ROM (c).}
	\label{fig:LSFS_r100_U384}
\end{figure}

The ability of the ROM to capture the large scale flow structures is now evaluated by comparing the vorticity and temperature fields for steady convection solutions with moderate $\R$ and $Pr=1$. In these cases, it is also interesting to compare the evolution of the amplitude coefficients during the transient towards the steady state solution.

For that, modes are first upsampled to match the DNS resolution in the $x$ direction. This upsampling enables the direct projection of DNS results onto the modal basis and is achieved via zero padding in the Fourier space. The DNS results are projected onto the modal basis to obtain the time evolution of the amplitude coefficients corresponding to the DNS solution: $a_i^\mathrm{DNS}$, $b_i^\mathrm{DNS}$ and $c_i^\mathrm{DNS}$. The equations of the ROM are integrated in time using the same random initial conditions as in DNS (cf. Tab.~\ref{tab:DNS}). In this section, the comparison between DNS and ROMs is done for models U384 and C384 with $n=384$ DoF.

Figure~\ref{fig:LSFS_r100_C384} compares steady state convection structures and the transient dynamics between DNS and model C384 at $\R=100$ (\CORR{$Ra=170\>777$}). In Fig.~\ref{fig:LSFS_r100_C384}a, the time evolution of the amplitude coefficients, $c_i$, over the initial transient is shown. Only the evolution of the five amplitude coefficients with the largest absolute values is represented in Fig.~\ref{fig:LSFS_r100_C384}. 
During the transient phase, the amplitude coefficients of the DNS and the ROM show some discrepancies but the overall trend is well captured.
 Figure~\ref{fig:LSFS_r100_C384}b and Fig.~\ref{fig:LSFS_r100_C384}c show the vorticity and temperature fields in DNS and ROM simulations when steady state is reached. 
 In steady state conditions, the reconstructed vorticity and temperature fields of the ROM show a  good agreement with the DNS. The flow structure consists of two steady counter-rotating convection rolls in the $x$ direction driving one vertical plume, with both the intensity and spatial distribution of vorticity and temperature fields accurately reproduced by the ROM. While the ROM captures the overall spatial distribution, small oscillations around the DNS solution can be observed in the 2D fields, resulting from the Gibbs phenomenon, as the low number of modes in the ROM does not allow a proper resolution of the sharp gradients in the DNS solution.

Finally, flow structures and transient dynamics are compared to DNS results in the uncoupled model U384. In this case, the solution fields for velocity and temperature are projected over the velocity and temperature modal bases, respectively, to give the DNS amplitude coefficients, $a_i^\mathrm{DNS}$ and $b_i^\mathrm{DNS}$. The results are shown in Fig.~\ref{fig:LSFS_r100_U384} for $\R=100$ ($Ra=170\>077$) and $Pr=1$. For the sake of clarity, only the five amplitude coefficients with the largest magnitude have been plotted in Fig.~\ref{fig:LSFS_r100_U384}a. Model U384 captures with great accuracy the transient phase towards the fixed point solution: ROM and DNS coefficients overlap and even the small oscillations occurring during the initial transient are reproduced by the ROM. In this case, the agreement with DNS in terms of vorticity and temperature fields is also very good and qualitatively comparable to that of the coupled ROM.

As illustrated in this section, the largest ROMs accurately reproduce steady-state flow features up to moderate reduced Rayleigh numbers. \CORR{Models U384 and C384 also capture the evolution of the amplitude coefficients during the initial growth and their steady saturation values. This indicates that the ROMs accurately reproduce the nonlinear energy transfers that balance the linear instability growth to establish the steady convection state.}

\subsection{Bifurcation diagram and power spectrum}\label{sec:bifurcation_analysis}

The preceding sections have demonstrated that both coupled and uncoupled ROMs accurately reproduce the mean vertical profiles, turbulence statistics, total heat flux, and large-scale flow structures in Rayleigh-Bénard convection. Beyond reproducing statistical quantities, a complete validation requires that these reduced-order systems also capture the dynamical regime of the system and its spectral characteristics. 
The route to chaos in Rayleigh-Bénard convection has been characterized in several numerical studies. For no-slip boundary conditions and $Pr = 0.71$, \citet{mclaughlin1982transition} analyzed the transition to chaos using 3D simulations and observed that periodic and quasiperiodic states precede the onset of chaos in the range of Rayleigh numbers between 6500 and $25\,000$. Similarly, in a 2D configuration with $Pr = 6.8$ and stress-free boundary conditions, \citet{paul2012bifurcation} observed that the system transitions from fixed-point solutions to chaos via periodic and quasiperiodic states, and at high reduced Rayleigh numbers returns to a fixed-point solution through a boundary crisis~\citep{grebogi1983crises}. 
These studies highlight the rich dynamical behavior of Rayleigh-Bénard convection and the importance of verifying that ROMs preserve the correct bifurcation sequence. In what follows, the analysis of bifurcation diagrams and power spectra is performed at $Pr = 10$ using $n = 192$ DoF ROMs.

\begin{figure}[!h]
	\centering
	\nolinenumbers
	\begin{subfigure}{0.32\textwidth}
		\centering
		\caption{DNS, $\R=40$}
		\includegraphics[width=1\textwidth]{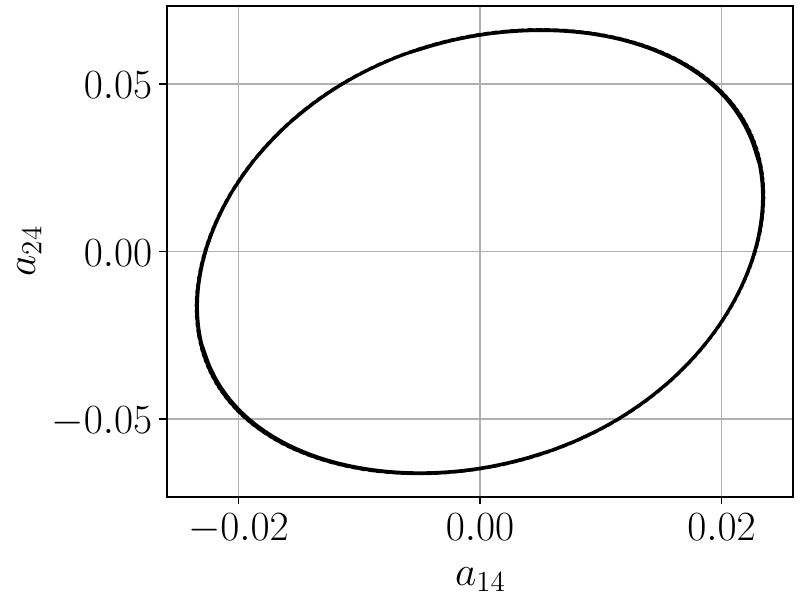}
		\label{fig:DNS_PP_r_040}
	\end{subfigure}
	\begin{subfigure}{0.32\textwidth}
		\centering
		\caption{DNS, $\R=80$}
		\includegraphics[width=1\textwidth]{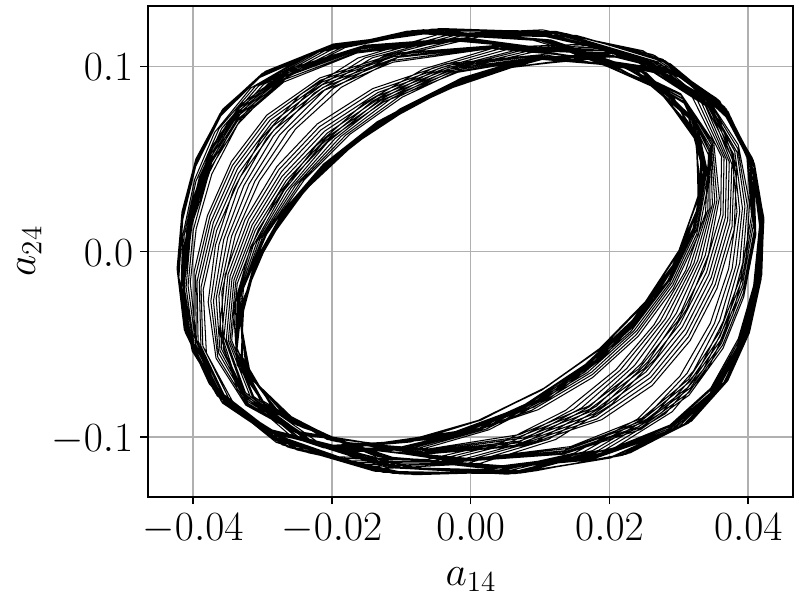}
		\label{fig:DNS_PP_r_080}
	\end{subfigure}
	\begin{subfigure}{0.32\textwidth}
		\centering
		\caption{DNS, $\R=120$}
		\includegraphics[width=1\textwidth]{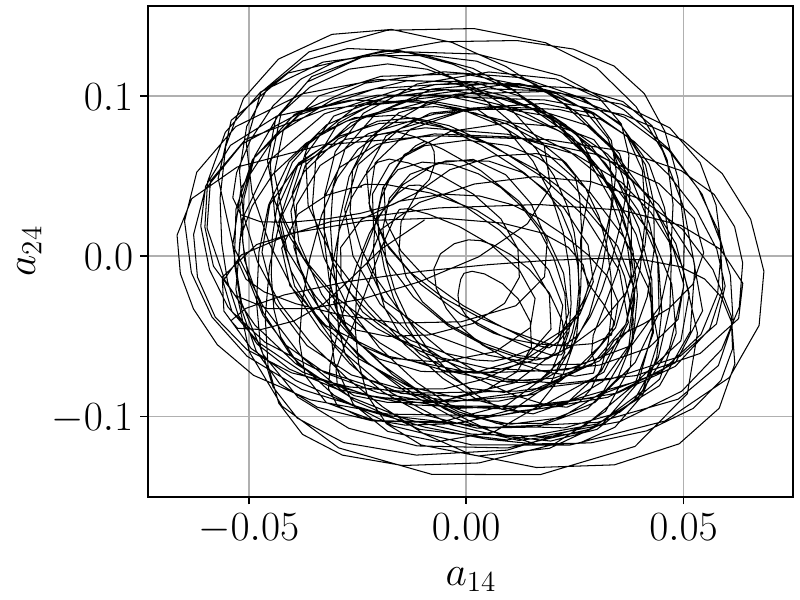}
		\label{fig:DNS_PP_r_120}
	\end{subfigure}
	\hfill
	\begin{subfigure}{0.32\textwidth}
		\centering
		\caption{C192, $\R=40$}
		\includegraphics[width=1\textwidth]{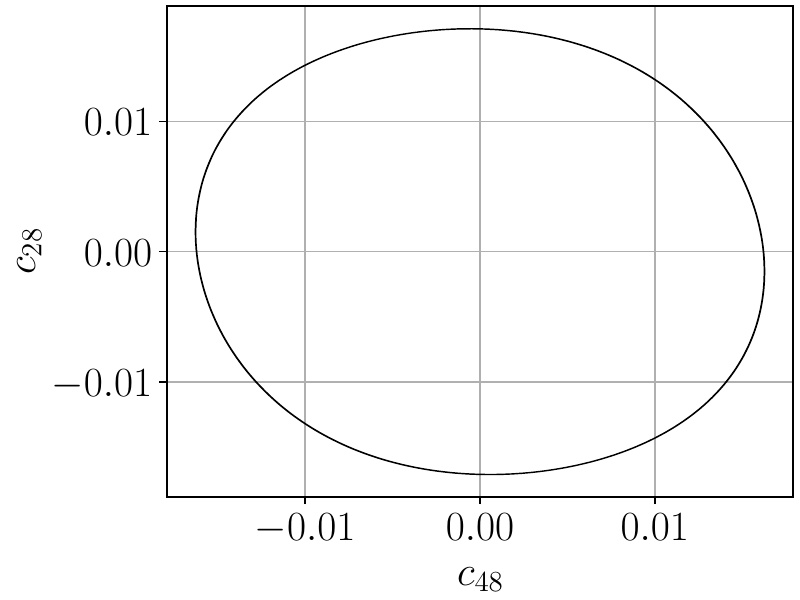}
		\label{fig:PhaPor_coupled_r040}
	\end{subfigure}
	\hfill
	\begin{subfigure}{0.32\textwidth}
		\centering
		\caption{C192, $\R=80$}
		\includegraphics[width=1\textwidth]{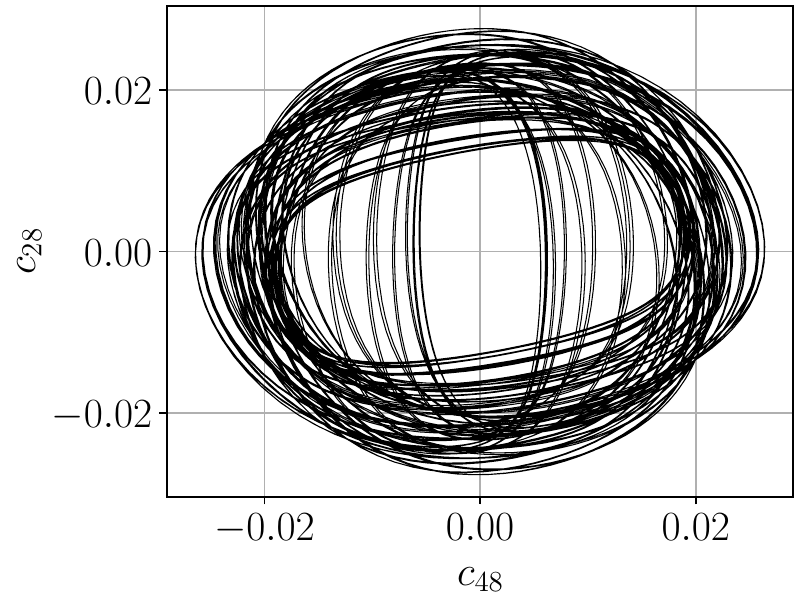}
		\label{fig:PhaPor_coupled_r080}
	\end{subfigure}
	\hfill
	\begin{subfigure}{0.32\textwidth}
		\centering
		\caption{C192, $\R=120$}
		\includegraphics[width=1\textwidth]{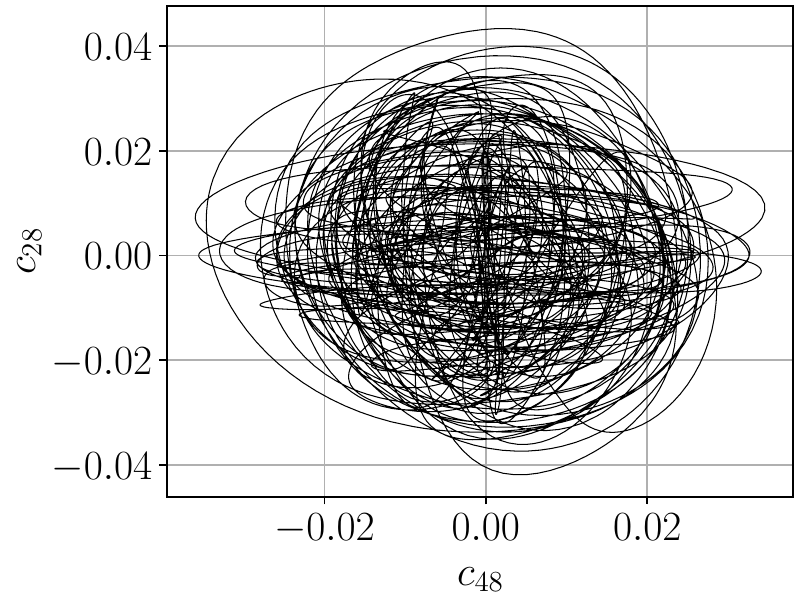}
		\label{fig:PhaPor_coupled_r120}
	\end{subfigure}
		\hfill
	
	\begin{subfigure}{0.32\textwidth}
		\centering
		\caption{U192, $\R=40$}
		\includegraphics[width=1\textwidth]{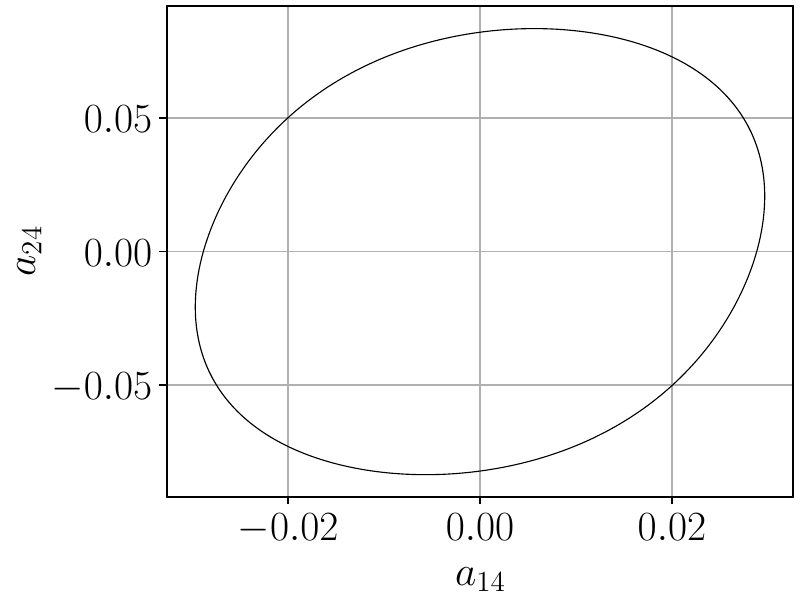}
		\label{fig:PhaPor_uncoupled_r040}
	\end{subfigure}
	\hfill
	\begin{subfigure}{0.32\textwidth}
		\centering
		\caption{U192, $\R=80$}
		\includegraphics[width=1\textwidth]{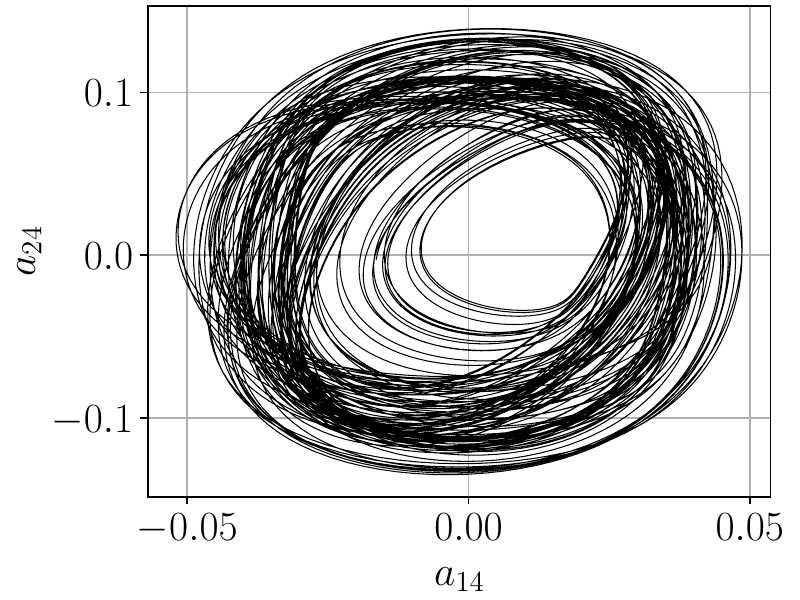}
		\label{fig:PhaPor_uncoupled_r080}
	\end{subfigure}
	\hfill
	\begin{subfigure}{0.32\textwidth}
		\centering
		\caption{U192, $\R=120$}
		\includegraphics[width=1\textwidth]{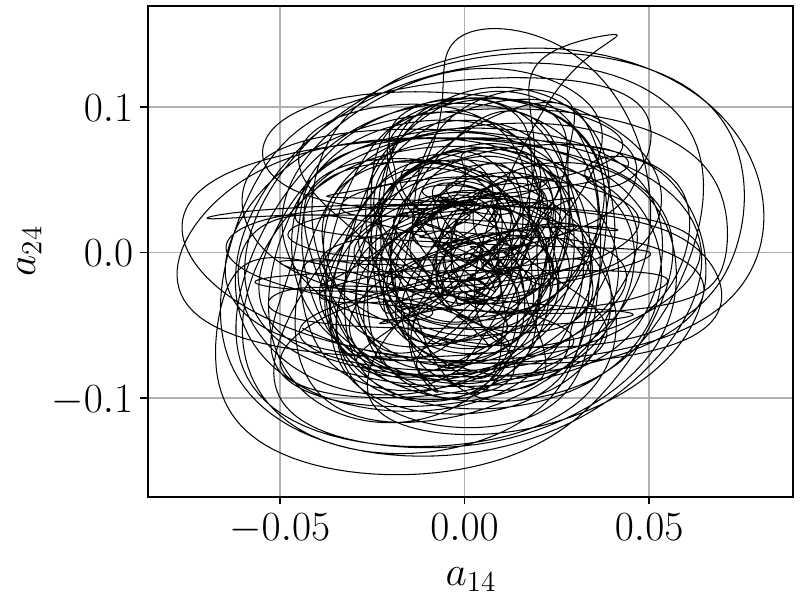}
		\label{fig:PhaPor_uncoupled_r120}
	\end{subfigure}
	\caption{\CORR{Phase portraits in DNS (a--c), model C192 (d--f) and model U192 (g--i) for different $\R$ numbers. DNS fields are projected onto the U192 model basis.}}
	\label{fig:phase_portraits}
\end{figure}

We begin by analyzing the route to chaos in DNS using phase portraits. DNS have been performed for several Rayleigh numbers at $Pr=10$, initialized using the velocity and temperature fields obtained with model U192 for each $\R$ number.  This reduces the transient duration and ensures that the state falls within the basin of the same attractor.
The equations have been integrated over a total time of $T=1500(Pr)^{-1/2}$, but only the second half of each simulation has been represented in the phase portraits. The DNS results have been projected onto the $n=192$ uncoupled ROM basis. For that, modes have been upsampled in the $x$ direction via zero-padding to match the DNS resolution. 

Figures~\ref{fig:DNS_PP_r_040}--\ref{fig:DNS_PP_r_120} show the DNS phase portraits at $\R=40$, 80, and 120, using the projection of DNS results onto modes $a_{14}$ and $a_{24}$. At $\R=40$, the phase portrait shows a stable orbit corresponding to periodic dynamics. At $\R=80$, the phase portrait exhibits quasiperiodic dynamics, characterized by a drift in the orbit trajectory over time. At $\R=120$, the system is chaotic and the trajectories densely fill the attractor region. DNS results indicate that for this Prandtl number, the system transitions to chaos via quasiperiodicity within the range $40<\R<120$.
 
 We now examine the dynamical regimes of coupled and uncoupled ROMs at the same Rayleigh numbers. \CORR{Figures~\ref{fig:PhaPor_coupled_r040}--\ref{fig:PhaPor_coupled_r120} show the phase portraits for coupled ROM C192 at the same Rayleigh numbers. Since DNS results were projected onto a different modal basis, the phase portraits of model C192 are constructed using modes $c_{28}$ and $c_{48}$. Therefore, the comparison focuses on the dynamical regime rather than on the direct agreement of the phase portraits with DNS results from Figs.~\ref{fig:DNS_PP_r_040}--\ref{fig:DNS_PP_r_120}. Model C192 displays periodic dynamics at $\R=40$. At $\R=80$, the system is quasiperiodic and the orbit amplitude increases relative to the $\R=40$ case. At $\R=120$, model C192 is chaotic and the phase portrait shows a dense cloud of trajectories in the $c_{28}$--$c_{48}$ plane.
 	
 Figures~\ref{fig:PhaPor_uncoupled_r040}--\ref{fig:PhaPor_uncoupled_r120} show the phase portraits for model U192 at $\R=40$, 80, and 120, analyzed using modes $a_{14}$ and $a_{24}$. At $\R=40$, the dynamics are periodic and modes $a_{14}$ and $a_{24}$ describe a stable orbit. Compared to the DNS projection, the orbit amplitude is slightly larger, but the attractor shape is very similar, indicating that the ROM reproduces the high-order system dynamics with good accuracy. At $\R=80$, the system exhibits quasiperiodic behavior, as evidenced by the drift in attractor orbits. At $\R=120$, the uncoupled ROM becomes chaotic, matching the DNS dynamical regime.
 
}
 
Models \CORR{C192 and U192} also transition to chaos via quasiperiodicity. Thus, they preserve the correct sequence of dynamical regimes in Rayleigh-Bénard convection at this Prandtl number, which is the essential requirement for using these models to investigate transition mechanisms. The detailed bifurcation analysis presented below demonstrates that the ROMs enable systematic characterization of these transitions at a fraction of the computational cost of DNS.

\CORR{To provide a quantitative estimate of the computational savings, we compared the wall-clock times of serial (single-core) simulations at $\R=120$ and $Pr=10$. The results were averaged over five simulations with different initial conditions. After discarding the initial transient, we timed the integration over $T=500$ time units. The DNS ($n_x=128$, $n_y=64$) took approximately $1155$~s, whereas the coupled ROMs required roughly $5$~s ($n=96$), $35$~s ($n=192$), and $285$~s ($n=384$), corresponding to speedups of approximately 225, 33, and 4, respectively. It should be noted that these figures represent a conservative baseline. While the DNS relies on Dedalus~\cite{burns2020dedalus}, a highly optimized spectral PDE framework, the current ROMs are integrated using a baseline Python implementation relying on general-purpose ODE integrators~\cite{hindmarsh1983odepack, petzold1983automatic} without dedicated performance optimization. Because this implementation was computationally sufficient for the present analysis, no further optimization was pursued. Even so, the unoptimized ROM provides substantial computational savings (up to two orders of magnitude for the most truncated models). Future work will focus on the development of tailored, high-performance numerical solvers to further optimize the integration of the ROM.}

\begin{figure}
	\centering
	\nolinenumbers
	\begin{subfigure}{0.32\textwidth}
		\centering
		\caption{C192, $\R=40$}
		\includegraphics[width=1\textwidth]{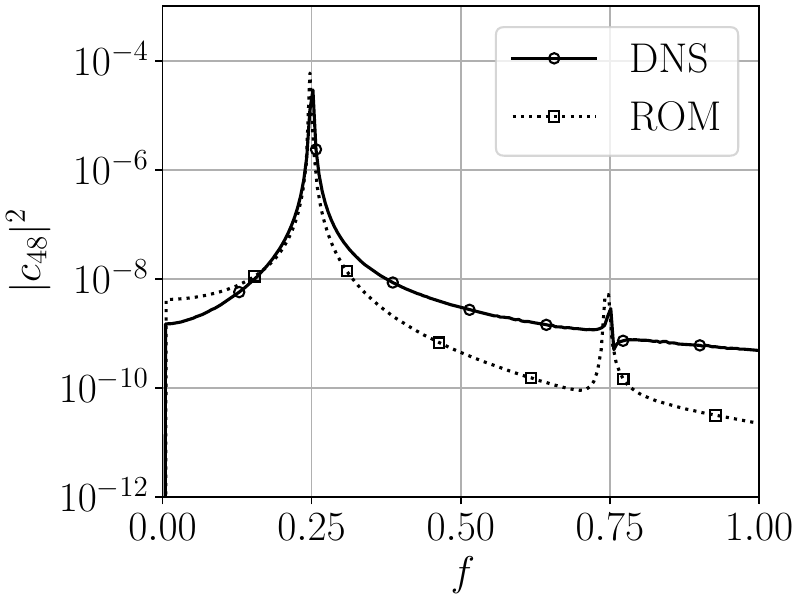}
		\label{fig:FFT_r040_C192}
	\end{subfigure}
	\begin{subfigure}{0.32\textwidth}
		\centering
		\caption{C192, $\R=80$}
		\includegraphics[width=1\textwidth]{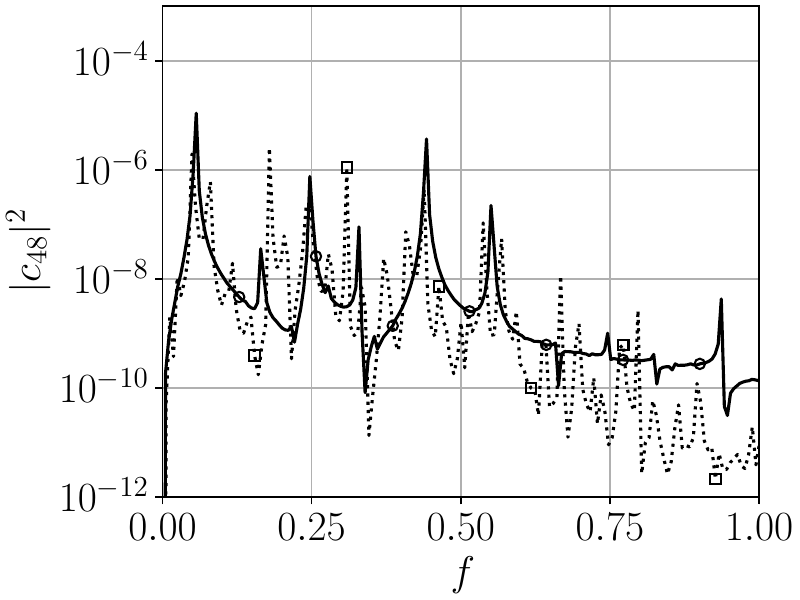}
		\label{fig:FFT_r080_C192}
	\end{subfigure}
	\begin{subfigure}{0.32\textwidth}
		\centering
		\caption{C192, $\R=120$}
		\includegraphics[width=1\textwidth]{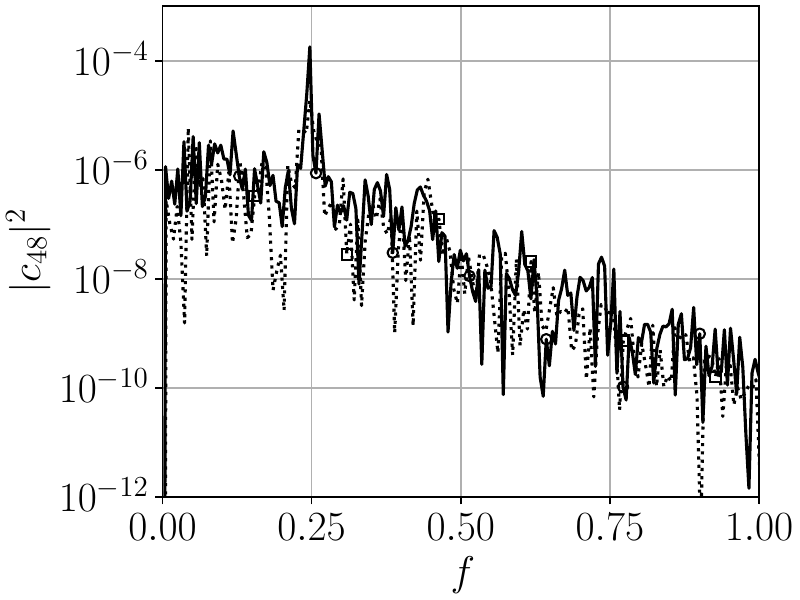}
		\label{fig:FFT_r120_C192}
	\end{subfigure}
	\hfill
	\begin{subfigure}{0.32\textwidth}
		\centering
		\caption{U192, $\R=40$}
		\includegraphics[width=1\textwidth]{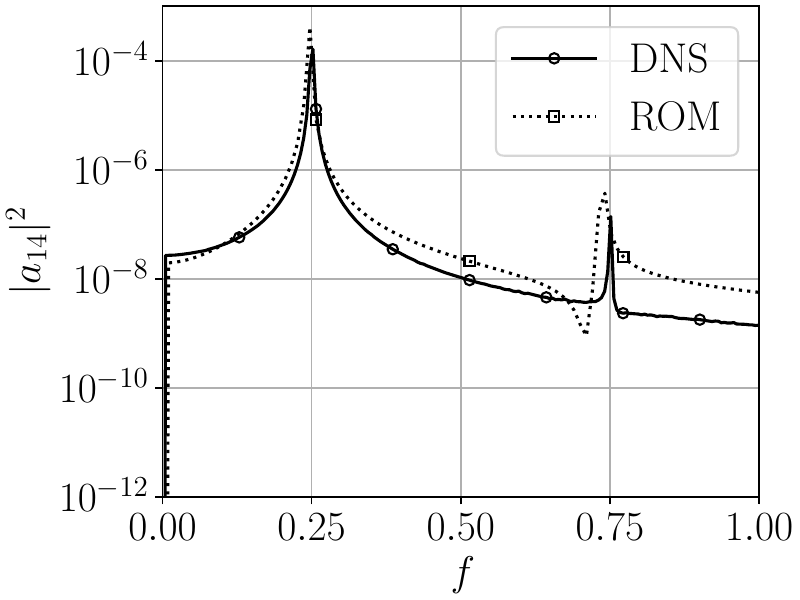}
		\label{fig:FFT_r040_U192}
	\end{subfigure}
	\begin{subfigure}{0.32\textwidth}
		\centering
		\caption{U192, $\R=80$}
		\includegraphics[width=1\textwidth]{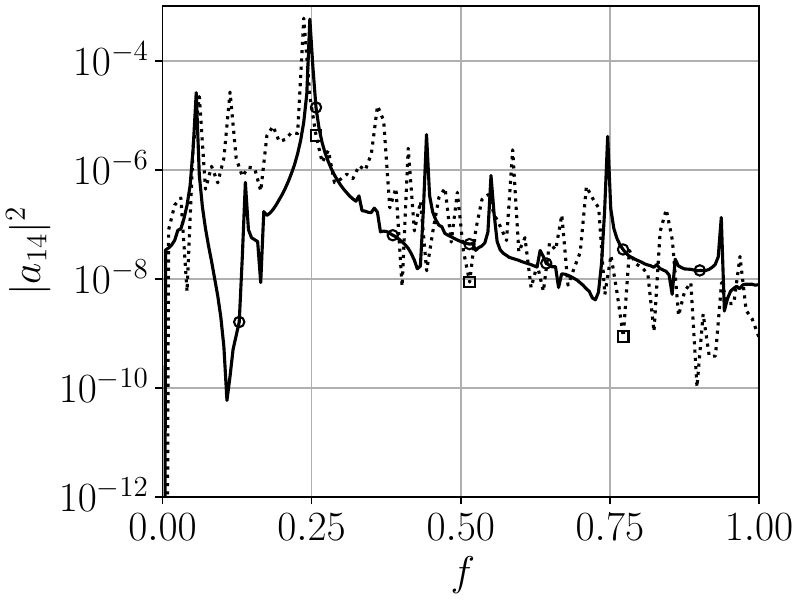}
		\label{fig:FFT_r080_U192}
	\end{subfigure}
	\begin{subfigure}{0.32\textwidth}
		\centering
		\caption{U192, $\R=120$}
		\includegraphics[width=1\textwidth]{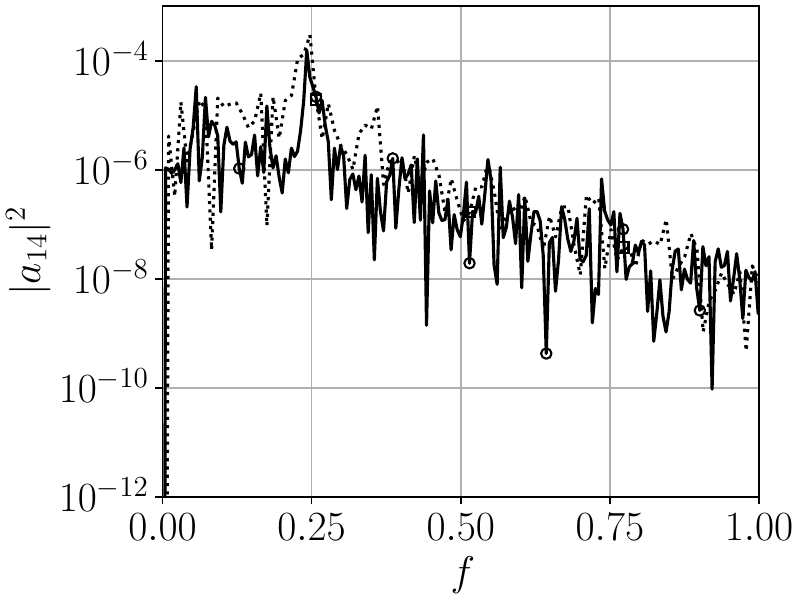}
		\label{fig:FFT_r120_U192}
	\end{subfigure}
	\hfill
	\caption{Power spectral density at different Rayleigh numbers and $Pr=10$: comparison between DNS and ROM results is established using mode $c_{48}$ for model C192 and mode $a_{14}$ for model U192.}
	\label{fig:power_spectrum}
\end{figure}

The projection of DNS results onto the orthonormal modal basis also enables a comparison of the power spectral density of the modal amplitude coefficients between DNS and ROM. In Fig.~\ref{fig:power_spectrum}, DNS and ROM power spectra are compared at $Pr=10$ for various Rayleigh numbers. DNS fields have been projected onto both the uncoupled and coupled modal bases for comparison. The spectra of modes $c_{48}$ and $a_{14}$ are used to analyze the frequency characteristics of models C192 and U192, respectively.

Figures~\ref{fig:FFT_r040_C192} and~\ref{fig:FFT_r040_U192} show the power spectra at $\R=40$. At this Rayleigh number, the system exhibits periodic behavior, and the power spectra display a single peak at $f\simeq0.25$. Excellent agreement is observed in both the frequency and amplitude of this spectral peak between DNS and ROM. The fact that ROMs not only capture the frequency but also the amplitude of the peak implies that they predict the amplitude of limit-cycle oscillations, thereby reproducing the nonlinear balance between destabilizing buoyancy and dissipation.

When the Rayleigh number is increased to $\R=80$, the system transitions to quasiperiodic behavior, and the power spectra exhibit multiple peaks at distinct frequencies. These results are shown in Figs.~\ref{fig:FFT_r080_C192} and~\ref{fig:FFT_r080_U192} for models C192 and U192, respectively. Model C192 captures the amplitude and frequency of the three dominant peaks in the DNS spectrum at $f\simeq0.05$, $0.25$, and $0.45$. Although the ROM spectrum contains two spurious peaks at $f\simeq0.16$ and $f\simeq0.32$, the overall agreement in terms of spectral energy distribution is good for model C192 at $\R=80$. Model U192 accurately reproduces the dominant spectral peak at $f\simeq0.25$, showing good agreement in both amplitude and frequency. This ROM also captures the amplitude and frequency of the low-frequency component at $f\simeq0.05$. However, the ROM spectrum exhibits two spurious peaks that do not correspond to frequency components present in the DNS signal. The spurious peaks observed in both ROM spectra are likely a consequence of modal truncation. Nevertheless, both models capture the dominant frequencies that characterize the quasiperiodic dynamics, indicating that the essential nonlinear interactions are preserved.

Finally, Figs.~\ref{fig:FFT_r120_C192} and~\ref{fig:FFT_r120_U192} show the power spectra at $\R=120$, where the system dynamics are chaotic. The DNS spectrum is broadband with a dominant peak at $f\simeq0.25$ and exhibits decay at higher frequencies. The power spectrum of mode $c_{48}$ in model C192 shows very good agreement with the DNS, capturing the amplitude and frequency of the dominant peak, the broadband spectral content, and the spectral decay rate at high frequencies. Model U192 also captures the dominant frequency within the broadband spectrum and the high-frequency decay rate. The $n=192$ ROMs reasonably reproduce the power spectrum of the system up to $\R=120$ while correctly representing the dynamical regimes---periodic, quasiperiodic, and chaotic---across the range of Rayleigh numbers investigated. Among these regimes, the quasiperiodic state proves most challenging to match in terms of spectral content, as it requires accurate prediction of multiple discrete frequencies rather than a single peak or a broadband distribution.

The preservation of the route to chaos in larger ROMs, combined with the good agreement in orbit amplitudes and phase portrait topology, indicates that these models accurately represent the fundamental physics of the system. This makes them valuable tools for flow control and state estimation applications, as well as for investigating transition mechanisms in Rayleigh-Bénard convection, \CORR{especially when computational efficiency is essential.}

\begin{figure}[!h]
     \centering
     \nolinenumbers
         \includegraphics[width=0.5\textwidth]{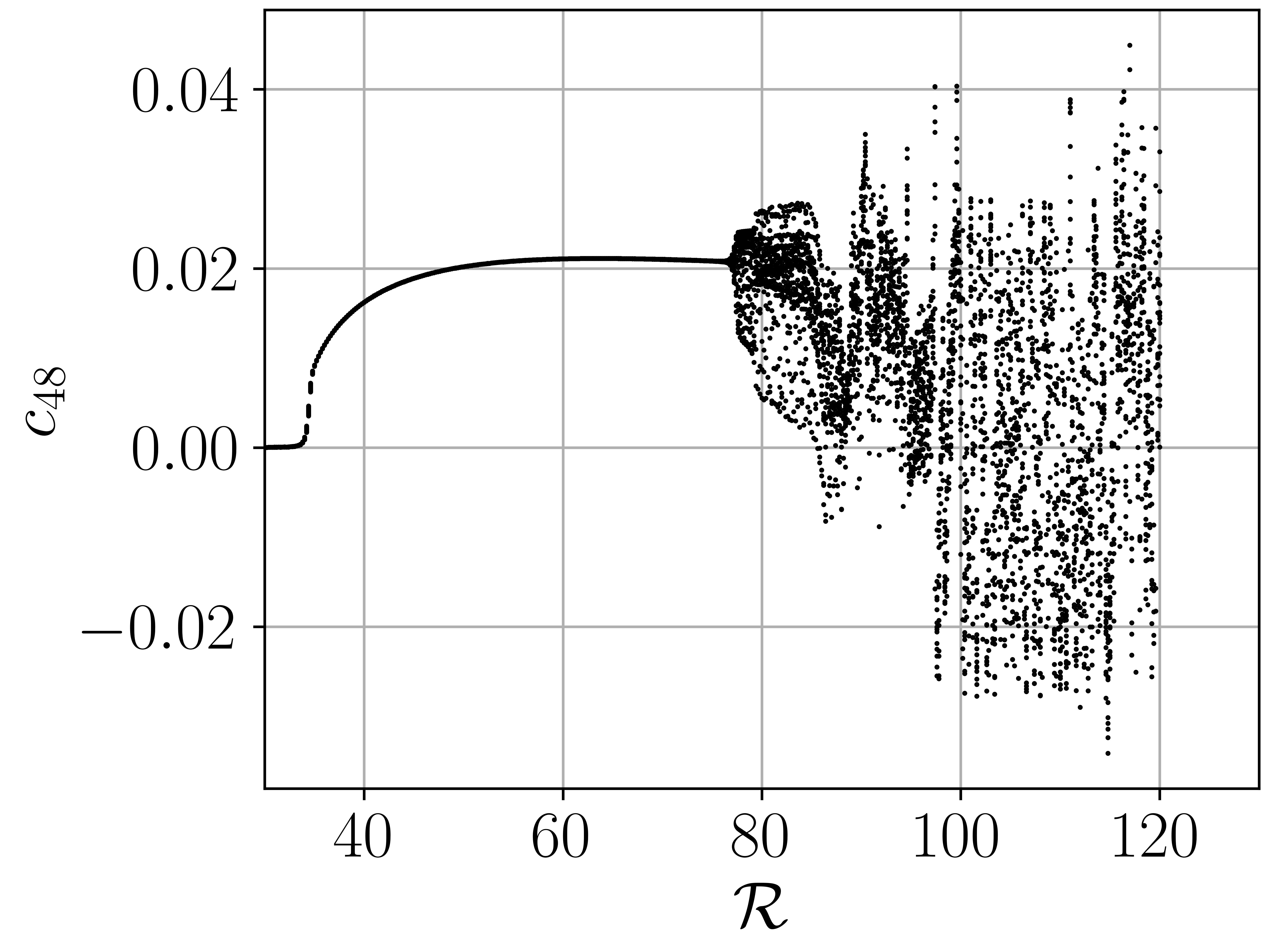}
        \caption{Poincaré section of model C192 \CORR{as a} function of the reduced Rayleigh number, $\R$.}
		\label{fig:poincare_coupled_n192}
\end{figure}

Having established the fidelity of the ROMs in reproducing the spectral characteristics of the flow, we now demonstrate the capabilities of these models by performing a detailed analysis of the route to chaos in the reduced order system. To this end, we combine Poincar\'{e} sections with the computation of Lyapunov exponents to precisely identify the transitions between different dynamical regimes. In what follows, we consider model C192, as the coupled basis approach has demonstrated better agreement with DNS in terms of Nusselt number and turbulence statistics.

The Poincaré section is used to visualize the system dynamics and is defined at the hyperplane $c_{28} = 0$. Only the crossings of the hyperplane in one direction, $\dot{c}_{28}>0$, are retained for the Poincaré section. This mapping allows us to track the evolution of the dynamical system. The analysis is performed by increasing the Rayleigh number in small steps, using the final state of each simulation as the initial condition for the subsequent one to minimize transients and ensure that the same solution branch is tracked. Each simulation runs for {$T=500Pr^{-1/2}$} time units, with only the final half of the data points being retained for the Poincaré section. This systematic approach reveals the attractor structure at each $Ra$: stable limit cycles appear as lines, quasiperiodic behavior produces dense regions, and chaos is characterized by scattered patterns. This approach has been used by \citet{kashinath2014nonlinear, cavalieri2022transition} in the analysis of \CORR{a} nonlinear thermoacoustic system and a shear layer, respectively. 

The Poincaré section of model C192 is shown in Fig.~\ref{fig:poincare_coupled_n192} using mode $c_{48}$. The system transitions from a fixed-point solution to periodic dynamics at approximately $\R\simeq34$. Then, the system transitions to quasiperiodicity at roughly $\R\simeq 75$, resulting in a dense region of crossing points in the Poincaré section. Finally, the system becomes chaotic at approximately $\R\simeq 88$, leading to a scattered pattern in the Poincaré section. However, the exact $\R$  number where the system transitions to chaos is hard to identify from the Poincaré section alone.  

\begin{figure}
	\centering
	\nolinenumbers
	\begin{subfigure}{0.48\textwidth}
		\centering
		\caption{Time evolution of \CORR{leading} Lyapunov exponents $\lambda_i(t)$.}
		\includegraphics[width=1\textwidth]{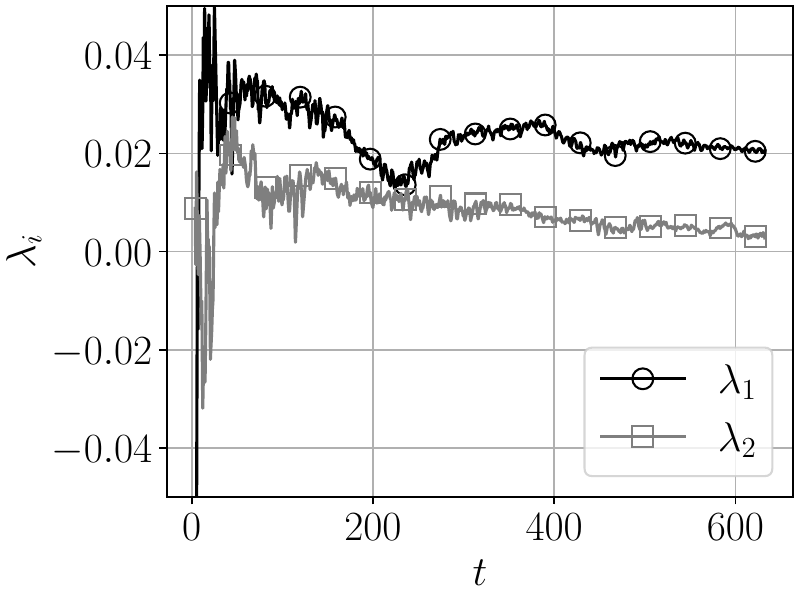}
		\label{fig:lyap_time}
	\end{subfigure}
	\hfill
	\begin{subfigure}{0.48\textwidth}
		\centering
		\caption{Variation of leading Lyapunov exponents with $\R$.}
		\includegraphics[width=1\textwidth]{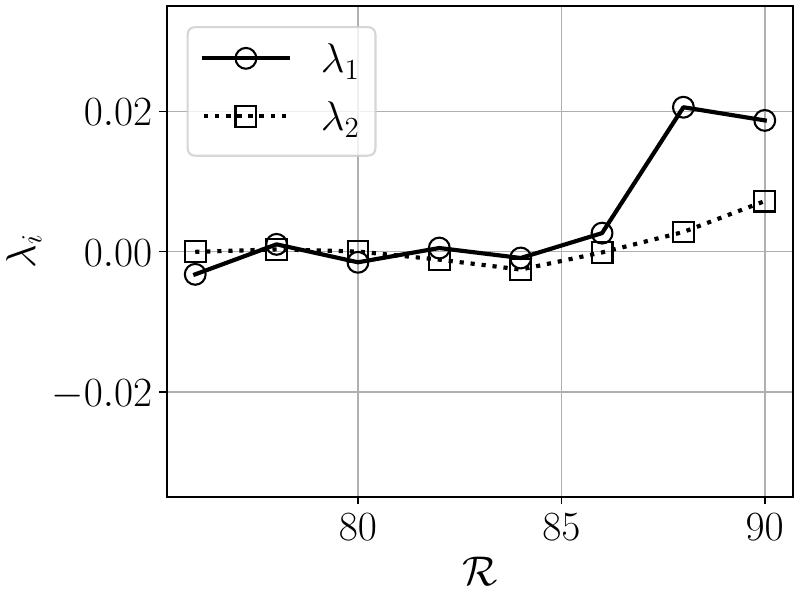}
		\label{fig:lyap_Ra}
	\end{subfigure}
	\caption{(a) Time evolution of the two leading Lyapunov exponents, $\lambda_i(t)$, in model C192 for $\R=88$ and $Pr=10$. (b) Variation of leading Lyapunov exponents with the $\R$ number at $Pr=10$ in model C192.}
	\label{fig:lyapunov}
\end{figure}

Since the leading Lyapunov exponent becomes positive when the system transitions to chaos, computing the leading Lyapunov exponents enables precise determination of the onset of chaotic dynamics. To this end, we have computed the two leading Lyapunov exponents using the continuous method described in~\citet{geist1990comparison} at different Rayleigh numbers close to the transition to chaos. In this method, the differential equations of the largest $k$ Lyapunov exponents are solved together with the system's equations. The dynamics of the Lyapunov exponents are described by the following differential equations,
\begin{eqnarray}\label{eq:Q_mat}
	\dot{Q} &=& JQ - QW,\\
\label{eq:lyap_dyn} \dot{\rho_i} &=& W_{ii}.
\end{eqnarray}
In eq.~\eqref{eq:Q_mat}, $Q$ is a $n\times k$ matrix, $J$ is the Jacobian matrix of the ROM and $W$ is a $k\times k$ upper triangular matrix defined as,
\begin{equation}\label{eq:Wmat}
	W_{ij} = \begin{cases}
		Q^TJQ + (Q^TJQ)^T + \nu  (Q^T Q - I) \quad&i<j,\\
		Q^TJQ + \nu  (Q^T Q - I)\quad&i=j,\\
		0\quad&i>j.
	\end{cases}
\end{equation}
In  eq.~\eqref{eq:Wmat}, \CORR{$\nu (Q^T Q - I)$} is a correction term to ensure the orthonormality of matrix $Q$, as during the integration of the differential equations, round-off errors can cause nonorthogonal perturbations to grow in time. The largest $k$ Lyapunov exponents are computed as $\lambda_i=\rho_i(t)/t$ in the limit $t\rightarrow\infty$. 

\CORR{To remove the influence of transients during the calculation of Lyapunov exponents}, the system is initialized from a trajectory on the attractor basin at each Rayleigh number. The QR decomposition of the Jacobian at $t=0$ is used to initialize the $Q$ matrix by taking the first $k$ columns of the orthogonal matrix. The initial condition for $\rho$ is $\rho_i(t=0)=0$. Equations~\eqref{eq:Q_mat} and~\eqref{eq:lyap_dyn} are integrated together with the dynamical system equations for $T=2000Pr^{-1/2}$, with the restoring force factor set to $\nu=1$. It has been verified that this continuous method yields values consistent with the discrete method based on QR decomposition described by~\citet{geist1990comparison} and~\citet{jolly2011numerical}.

Figure~\ref{fig:lyap_time} shows the time evolution of the two leading Lyapunov exponents, $\lambda_1$ and $\lambda_2$, in system C192 for $\R=88$. Both Lyapunov exponents are positive, confirming that the system dynamics are chaotic at this Rayleigh number. Figure~\ref{fig:lyap_Ra} shows the evolution of the two leading Lyapunov exponents with Rayleigh number for model C192. For each Rayleigh number, the final value of $\lambda_i$ from each simulation is shown. When the system is quasiperiodic, the two leading Lyapunov exponents are zero, indicating that the distance between adjacent trajectories neither grows nor decays with time. When one Lyapunov exponent becomes positive, the distance between nearby trajectories grows exponentially. The results presented in Fig.~\ref{fig:lyap_Ra} show that system C192 transitions to chaos at $\R\simeq87$.

The combination of Poincar\'{e} sections and Lyapunov exponent analysis presented here illustrates a few key functionalities of these  reduced-order models. Computing Poincar\'{e} sections in DNS requires extensive integration times, while Lyapunov exponent calculation additionally requires the system Jacobian, which is straightforward to obtain in the ROM but computationally demanding in high-dimensional systems. Such a detailed bifurcation analysis would therefore be difficult to perform directly on DNS. This example demonstrates one potential application of these ROMs: the detailed investigation of transition mechanisms in nonlinear fluid systems. 

The comprehensive validation performed throughout this \CORR{work}---encompassing mean profiles, heat flux, turbulence statistics, spectral characteristics, and dynamical regimes---suggests that these models faithfully capture the physical behavior of the full system. This opens the door for their application in flow control and state estimation, where both computational efficiency and accuracy are essential.  

\section{Conclusions}\label{sec:conclusions}
Galerkin projection has been used to obtain different ROMs for two-dimensional Rayleigh-Bénard convection with no-slip walls. The methodology developed by~\citet{cavalieri2022reduced} to obtain orthonormal bases from the eigenvectors of the controllability Gramian has been {extended to the Boussinesq-Navier-Stokes system} to obtain orthonormal bases with no-slip boundary conditions, which are more amenable to experimental validation than stress-free walls. Two projection strategies have been explored: an uncoupled ROM with two independent bases for velocity and temperature, and a coupled ROM with one single orthonormal basis that includes velocity and temperature components. The coupled approach has required the derivation of the adjoint linearized equations for two-dimensional Rayleigh-Bénard convection. 
{The adjoint equations have been obtained using an inner product that weights the energy of the velocity and temperature fields.}

Coupled and uncoupled ROMs have been compared to two-dimensional DNS results in various ways, including mean vertical profiles of temperature, heat flux and Reynolds stresses,  Nusselt number, large-scale flow structures, steady-convection solutions, bifurcation diagrams and energy spectra. The validation has been performed for different modal truncations over a wide range of $Ra$ numbers. 
As expected, the agreement with DNS results in terms of mean vertical profiles and $Nu$ number increases with the ROM dimension and decreases with the $\R$ number. Nevertheless, our results show that a quantitative agreement with DNS can be obtained using coupled and uncoupled-basis ROMs with a few hundreds of modes over a wide range of $\R$ numbers. In this respect, the coupled projection strategy has been found to outperform the more conventional uncoupled approach suggesting that coupled modal bases better capture the nonlinear interactions between large scale structures. Furthermore, these ROMs have proven to be numerically stable when integrated over large characteristic times, $T=10^4,$ at high Rayleigh numbers, $\R= 10^6$. The direct comparison between steady DNS solutions {representing convection rolls and the corresponding ROM} results has shown that models U384 and C384 can reproduce the vorticity and temperature fields with quantitative agreement up to $\R=100$. In addition, these ROMs accurately capture the dynamics of the system, predicting the temporal evolution of the amplitude coefficients during the initial growth.

Finally, a detailed analysis of the dynamical behavior of the system as a function of $\R$ has been conducted for models \CORR{C192 and U192} at $Pr=10$. The dynamical regimes of the system have been studied using Poincar\'{e} sections, phase portraits, power spectra, and Lyapunov exponents. This bifurcation analysis has shown that these ROMs reproduce the same route to chaos observed in DNS---periodic, quasiperiodic, and chaotic dynamics---as a function of the Rayleigh number. Good agreement has been obtained for the phase portraits and power spectra up to $\R=120$. In particular, these ROMs capture both the frequency and amplitude of the motion for periodic solutions ($\R=40$) and the broadband frequency spectrum in chaotic states ($\R=120$). The quasiperiodic regime ($\R=80$) proves most challenging to match spectrally due to the presence of multiple discrete frequencies, though the dominant spectral components are correctly reproduced. The combination of Poincar\'{e} sections and Lyapunov exponent analysis enabled precise identification of the transition to chaos at $\R\simeq87$ in model C192, illustrating the utility of these ROMs for detailed bifurcation studies that would be computationally demanding to perform directly on DNS.

Overall, these results indicate that Galerkin ROMs reproduce the main features of the nonlinear system dynamics and are promising candidates to serve as surrogates for the high-dimensional system in applications where a low computational cost is required. This includes the derivation of optimal control strategies, state estimation from local measurements and the fundamental study of system bifurcations using dynamical system tools. Future works will explore the use of these ROMs in control and state estimation applications for RB convection. Other research directions include the extension to three dimensions and the incorporation of additional equations to account for species-driven transport~\cite{turton2015prediction}.

\CORR{As a direct application of the present findings, our companion paper~\cite{flores2026state} demonstrates the practical use of these ROMs for flow state estimation. In~\cite{flores2026state}, coupled ROMs  are combined with an extended Kalman filter to reconstruct the flow fields from sparse measurements across periodic, quasiperiodic, and chaotic regimes.}

\section*{Acknowledgments}

A.V.G.C. acknowledges support from CNPq through grant 314927/2023-9.

{\paragraph*{Data availability statement.} The computational tools used in this study, including Python scripts for modal basis generation and ROM simulation, are made available as an open-source package at \url{https://github.com/kikeflores96/rayleigh_benard_rom.git}. The rest of the data and codes used to generate the results of the present work can be provided upon reasonable request to the authors.}

\appendix

\section{Mathematical developments}\label{sec:appendix_derivation}
In this Appendix, the mathematical developments required to obtain the expression for the inner product in the state space and for the state and control matrices of the adjoint system are presented.
\subsection{Inner product in the state space}
The derivation of the expression for the inner product in the state space begins with eq.~\eqref{eq:inner_product_SS},
\allowdisplaybreaks
\begin{align}
	\langle\boldsymbol{y}_1,\>W \boldsymbol{y}_2\rangle &=\frac{1}{L_y}\int_{0}^{L_y}\boldsymbol{y}_1^HW\boldsymbol{y}_2\,dy\nonumber\\
	&= \int_{0}^{1}\left(C\boldsymbol{z}_1\right)^H W C\boldsymbol{z}_2\,dy \nonumber\\
	&= \int_{0}^{1}\left(\begin{bmatrix}\partial_y & 0\\-ik_x & 0\\ 0& 1\end{bmatrix}\begin{bmatrix}\tilde{\Psi}_1\\ \tilde{\theta}_1\end{bmatrix}\right)^H \begin{bmatrix}1&0&0\\0&1&0\\0&0&\gamma^2  \end{bmatrix}  \begin{bmatrix}\partial_y & 0\\-ik_x & 0\\ 0& 1\end{bmatrix}\begin{bmatrix}\tilde{\Psi}_2\\ \tilde{\theta}_2\end{bmatrix} dy \nonumber\\
	&= \int_{0}^{1}\begin{bmatrix}\partial_y\tilde{\Psi}_1\\-ik_x\tilde{\Psi}_1\\ \tilde{\theta}_1\end{bmatrix}^H\begin{bmatrix}\partial_y\tilde{\Psi}_2\\-ik_x\tilde{\Psi}_2\\ \gamma^2 \tilde{\theta}_2\end{bmatrix} dy \nonumber\\
	&= \int_{0}^{1}\begin{bmatrix}\partial_y\tilde{\Psi}^\star_1 & ik_x\tilde{\Psi}^\star_1& \tilde{\theta}^\star_1\end{bmatrix}\begin{bmatrix}\partial_y\tilde{\Psi}_2\\-ik_x\tilde{\Psi}_2\\ \gamma^2 \tilde{\theta}_2\end{bmatrix} dy\nonumber\\
	&= \int_{0}^{1}\left(\partial_y\tilde{\Psi}^\star_1\partial_y\tilde{\Psi}_2 +k_x^2\tilde{\Psi}^\star_1\tilde{\Psi}_2  + \gamma^2\tilde{\theta}^\star_1\tilde{\theta}_2\right) dy,
\end{align}
where the decomposition into Fourier modes in the $x$ direction allows us to reduce the inner product to an integral in the $y$ direction only.
Applying integration by parts to the first term of the integrand and invoking the boundary conditions $\tilde{\Psi}(y=0)=\tilde{\Psi}(y=1)=0$, we obtain,
\begin{eqnarray}
    \nonumber\langle\boldsymbol{y}_1,\>W \boldsymbol{y}_2\rangle
    &=& \cancel{\left[\tilde{\Psi}^\star_1\partial_y\tilde{\Psi}_2\right]_0^1} + \int_{0}^{1}\left(-\tilde{\Psi}^\star_1\partial_y^2\tilde{\Psi}_2 + k_x^2\tilde{\Psi}^\star_1\tilde{\Psi}_2  + \gamma^2\tilde{\theta}^\star_1\tilde{\theta}_2\right) dy \\
    \nonumber&=& \int_{0}^{1}\left(\tilde{\Psi}^\star_1(- \nabla^2) \tilde{\Psi}_2 +\gamma^2\tilde{\theta}^\star_1\tilde{\theta}_2\right) dy \\
    \nonumber&=& \int_{0}^{1} \begin{bmatrix}   \tilde{\Psi}^\star_1& \tilde{\theta}^\star_1\end{bmatrix}\begin{bmatrix}
        -\nabla^2&0\\0&\gamma^2\end{bmatrix}\begin{bmatrix}        \tilde{\Psi}_2\\ \tilde{\theta}_2\end{bmatrix} dy \\
    &=&\int_{0}^{1}\boldsymbol{z}_1^HM\boldsymbol{z}_2dy,
\end{eqnarray}
with a weight matrix for the inner product given by,
\begin{equation}
    M=  \begin{bmatrix}-\nabla^2&0\\0& \gamma^2    \end{bmatrix},
\end{equation}
where $\nabla^2 = \partial_y^2 - k_x^2$ represents the two-dimensional Laplacian operator.
\subsection{Adjoint system matrices}
Firstly, we consider the adjoint state matrix, $A^+$, which is defined by,
\begin{equation}
    \langle A\boldsymbol{z}_1, \boldsymbol{z}_2 \rangle_e = \langle \boldsymbol{z}_1, A^+\boldsymbol{z}_2 \rangle_e,
\end{equation}
with an inner product in the state space, $\langle\cdot\rangle_e$ defined in eq.~\eqref{eq:inner_prod}. Expanding both sides of this equation gives,
\begin{equation}
     \int_0^1 (A\boldsymbol{z}_1)^H M \boldsymbol{z}_2 \, dy = \int_0^1 \boldsymbol{z}_1^H M A^+ \boldsymbol{z}_2 \, dy.
\end{equation}
Substituting the expressions for $A$ and $M$ from eq.~\eqref{eq:direct_state_matrix} and eq.~\eqref{eq:state_inner_prod} we obtain,
\begin{eqnarray}
     \nonumber\int_0^1 \left(\begin{bmatrix}
        \frac{Pr}{\sqrt{Ra}}\nabla^{-2}\nabla^4& -ik_xPr\nabla^{-2}\\ ik_x\mathrm{d}_y\theta_0 &\frac{1}{\sqrt{Ra}}\nabla^2
    \end{bmatrix}\begin{bmatrix} \tilde{\Psi}_1 \\ \tilde{\theta}_1 \end{bmatrix}\right)^H \begin{bmatrix} -\nabla^2 & 0 \\ 0 & \gamma^2 \end{bmatrix} \begin{bmatrix} \tilde{\Psi}_2 \\ \tilde{\theta}_2 \end{bmatrix} \, dy =\\
    \int_0^1 \begin{bmatrix} \tilde{\Psi}_1 \\ \tilde{\theta}_1 \end{bmatrix}^H \begin{bmatrix} -\nabla^2 & 0 \\ 0 & \gamma^2 \end{bmatrix} \begin{bmatrix} A_{11}^+ & A_{12}^+\\ A_{21}^+ & A_{22}^+  \end{bmatrix} \begin{bmatrix} \tilde{\Psi}_2 \\ \tilde{\theta}_2 \end{bmatrix} \, dy.
\end{eqnarray}
Computing the matrix products gives,
\begin{eqnarray}
     \nonumber\int_0^1 \begin{bmatrix}
        \frac{Pr}{\sqrt{Ra}}\nabla^{-2}\nabla^4\tilde{\Psi}^\star_1  +ik_xPr\nabla^{-2}\tilde{\theta}^\star_1\\ -ik_x\mathrm{d}_y\theta_0 \tilde{\Psi}^\star_1 + \frac{1}{\sqrt{Ra}}\nabla^2\tilde{\theta}^\star_1
    \end{bmatrix}^T \begin{bmatrix} -\nabla^2\tilde{\Psi}_2 \\ \gamma^2\tilde{\theta}_2 \end{bmatrix} \, dy =\\
    \int_0^1 \begin{bmatrix} \tilde{\Psi}^\star_1 & \tilde{\theta}^\star_1 \end{bmatrix} \begin{bmatrix} -\nabla^2A_{11}^+ \tilde{\Psi}_2  -\nabla^2A_{12}^+\tilde{\theta}_2\\ \gamma^2A_{21}^+\tilde{\Psi}_2 + \gamma^2A_{22}^+\tilde{\theta}_2  \end{bmatrix} \,dy .
\end{eqnarray}
Expanding this equation yields,
\begin{eqnarray}
     &&\nonumber\int_0^1 \left(-\frac{Pr}{\sqrt{Ra}}\nabla^{-2}\nabla^4\tilde{\Psi}^\star_1\nabla^2\tilde{\Psi}_2 - ik_xPr\nabla^{-2}\tilde{\theta}^\star_1 \nabla^2\tilde{\Psi}_2 -ik_x \gamma^2\mathrm{d}_y\theta_0 \tilde{\Psi}^\star_1\tilde{\theta}_2 + \frac{ \gamma^2}{\sqrt{Ra}}\nabla^2\tilde{\theta}^\star_1\tilde{\theta}_2\right)dy\\
    &=&\int_0^1 \left( -\tilde{\Psi}^\star_1\nabla^2A_{11}^+ \tilde{\Psi}_2 - \tilde{\Psi}^\star_1 \nabla^2A_{12}^+\tilde{\theta}_2 + \tilde{\theta}^\star_1\gamma^2A_{21}^+\tilde{\Psi}_2 + \tilde{\theta}^\star_1 \gamma^2A_{22}^+\tilde{\theta}_2 \right) \,dy.
\end{eqnarray}
Equating corresponding terms on both sides gives,
\begin{eqnarray}
    \label{eq:A11_ad_stp1}\int_0^1 \left(-\frac{Pr}{\sqrt{Ra}}\nabla^{-2}\nabla^4\tilde{\Psi}^\star_1\nabla^2\tilde{\Psi}_2\right) dy = \int_0^1 \left( -\tilde{\Psi}^\star_1\nabla^2A_{11}^+ \tilde{\Psi}_2\right) dy,\\
    \label{eq:A21_ad_stp1}\int_0^1 \left(- ik_xPr\nabla^{-2}\tilde{\theta}^\star_1 \nabla^2\tilde{\Psi}_2\right) dy = \int_0^1 \left( \tilde{\theta}^\star_1\gamma^2A_{21}^+\tilde{\Psi}_2\right) dy,\\
    \label{eq:A12_ad_stp1}\int_0^1 \left(-ik_x \gamma^2\mathrm{d}_y\theta_0 \tilde{\Psi}^\star_1\tilde{\theta}_2\right) dy = \int_0^1 \left(- \tilde{\Psi}^\star_1 \nabla^2A_{12}^+\tilde{\theta}_2\right) dy,\\
    \label{eq:A22_ad_stp1}\int_0^1 \left(\frac{ \gamma^2}{\sqrt{Ra}}\nabla^2\tilde{\theta}^\star_1\tilde{\theta}_2\right) dy = \int_0^1 \left(\tilde{\theta}^\star_1 \gamma^2A_{22}^+\tilde{\theta}_2\right) dy.
\end{eqnarray}
The Laplacian is \CORR{a} Hermitian operator meaning that for any two functions $\phi(y)$ and $\psi(y)$ satisfying $\phi|_{y=0,y=1}=0$ \CORR{and} $\psi|_{y=0,y=1}=0$ we have,
\begin{eqnarray}\label{eq:lapl_herm}
    \int_0^1\phi\nabla^2 \psi dy&=&\int_0^1\nabla^2\phi\psi dy.
\end{eqnarray}
In eq.~\eqref{eq:lapl_herm}, integration by parts and boundary conditions are used to transfer the Laplacian operator from one function to the other.
Applying this property to rewrite the left hand side of eqs.~\eqref{eq:A11_ad_stp1}--\eqref{eq:A22_ad_stp1} we obtain,
\begin{eqnarray}
    \label{eq:A11_ad_stp2}\int_0^1 \left(\tilde{\Psi}^\star_1\frac{Pr}{\sqrt{Ra}}\nabla^4\tilde{\Psi}_2\right) dy = \int_0^1 \left(\tilde{\Psi}^\star_1\nabla^2A_{11}^+ \tilde{\Psi}_2\right) dy,\\
    \label{eq:A21_ad_stp2}\int_0^1 \left(\tilde{\theta}^\star_1 (- ik_xPr)\tilde{\Psi}_2\right) dy = \int_0^1 \left( \tilde{\theta}^\star_1\gamma^2A_{21}^+\tilde{\Psi}_2\right) dy,\\
    \label{eq:A12_ad_stp2}\int_0^1 \left( \tilde{\Psi}^\star_1(ik_x \gamma^2\mathrm{d}_y\theta_0)\tilde{\theta}_2\right) dy = \int_0^1 \left( \tilde{\Psi}^\star_1 \nabla^2A_{12}^+\tilde{\theta}_2\right) dy,\\
    \label{eq:A22_ad_stp2}\int_0^1 \left(\tilde{\theta}^\star_1\frac{ 1}{\sqrt{Ra}}\nabla^2\tilde{\theta}_2\right) dy = \int_0^1 \left(\tilde{\theta}^\star_1 A_{22}^+\tilde{\theta}_2\right) dy.
\end{eqnarray}
Equating corresponding terms yields the expressions for the elements of the adjoint matrix,
\begin{eqnarray}
    \label{eq:A11_ad_stp3}\frac{Pr}{\sqrt{Ra}}\nabla^4 = \nabla^2A_{11}^+ &\rightarrow& A_{11}^+ = \frac{Pr}{\sqrt{Ra}}\nabla^{-2}\nabla^4,\\
    \label{eq:A21_ad_stp3} - ik_xPr=\gamma^2A_{21}^+&\rightarrow& A_{21}^+ = -ik_xPr/\gamma^2,\\
    \label{eq:A12_ad_stp3}ik_x \gamma^2\mathrm{d}_y\theta_0=  \nabla^2A_{12}^+ &\rightarrow& A_{12}^+= ik_x\gamma^2\nabla^{-2}\mathrm{d}_y\theta_0,\\
    \label{eq:A22_ad_stp3}\frac{ 1}{\sqrt{Ra}}\nabla^2 =  A_{22}^+ &\rightarrow& A_{22}^+  = \frac{ 1}{\sqrt{Ra}}\nabla^2,
\end{eqnarray}
which can now be arranged in matrix form to yield,
\begin{equation}
    A^+ = \begin{bmatrix} \frac{Pr}{\sqrt{Ra}}\nabla^{-2}\nabla^4 & ik_x\gamma^2\nabla^{-2}\mathrm{d}_y\theta_0 \\ -ik_xPr/\gamma^2 & \frac{1}{\sqrt{Ra}}\nabla^2 \end{bmatrix}.
\end{equation}
A similar procedure is now applied to obtain the expression of the adjoint control matrix, $B^+$, which is defined by,
\begin{equation}
\langle \boldsymbol{z}, B\boldsymbol{w} \rangle_e = \langle B^+\boldsymbol{z}, \boldsymbol{w} \rangle_{L^2}.
\end{equation}
Expanding the equation gives,
\begin{equation}
     \int_0^1 \boldsymbol{z}^H MB \boldsymbol{w} \, dy = \int_0^1 (B^+\boldsymbol{z})^H \boldsymbol{w} \, dy.
\end{equation}
Substituting the expressions for $B$ and $M$ we obtain,
\begin{eqnarray}
     \nonumber\int_0^1 \begin{bmatrix} \tilde{\Psi} \\ \tilde{\theta} \end{bmatrix}^H  \begin{bmatrix} -\nabla^2 & 0 \\ 0 & \gamma^2 \end{bmatrix} \begin{bmatrix}
        \nabla^{-2}\partial_y &-ik_x\nabla^{-2} & 0\\
        0& 0&1
    \end{bmatrix} \begin{bmatrix}\tilde{d}_x\\ \tilde{d}_y \\ \tilde{q} \end{bmatrix} \, dy =\\
    \int_0^1 \left(\begin{bmatrix}B^+_{11} &0 \\ B^+_{21} &0 \\0 &B^+_{32} \end{bmatrix} \begin{bmatrix} \tilde{\Psi} \\ \tilde{\theta} \end{bmatrix}\right)^H \begin{bmatrix}\tilde{d}_x\\ \tilde{d}_y \\ \tilde{q} \end{bmatrix} \, dy.
\end{eqnarray}
Computing the matrix products yields,
\begin{eqnarray}
     \nonumber\int_0^1 \left(- \tilde{\Psi}^\star\partial_y \tilde{d}_x + ik_x \tilde{\Psi}^\star\tilde{d}_y + \gamma^2\tilde{\theta}^\star\tilde{q}\right) \, dy =  \\
     = \int_0^1 \left( B_{11}^{+^\star}\tilde{\Psi}^\star\tilde{d}_x + B_{21}^{+^\star} \tilde{\Psi}^\star\tilde{d}_y +B_{32}^{+^\star}\tilde{\theta}^\star\tilde{q}\right) \, dy.
\end{eqnarray}
Applying integration by parts to the first term on the left-hand side and noting that the boundary term vanishes due to the boundary conditions, we obtain,
\begin{eqnarray}
     \nonumber-\cancel{\left[\tilde{\Psi}^\star\tilde{d_x}\right]_0^1 }+ \int_0^1 \left(\partial_y \tilde{\Psi}^\star\tilde{d}_x + ik_x \tilde{\Psi}^\star\tilde{d}_y + \gamma^2\tilde{\theta}^\star\tilde{q}\right) \, dy= \\
     = \int_0^1 \left( B_{11}^{+^\star}\tilde{\Psi}^\star\tilde{d}_x + B_{21}^{+^\star} \tilde{\Psi}^\star\tilde{d}_y +B_{32}^{+^\star}\tilde{\theta}^\star\tilde{q}\right) \, dy.
\end{eqnarray}
Equating corresponding terms on both sides of the equation gives,
\begin{eqnarray}
    \partial_y= B_{11}^{+^\star} &\rightarrow& B_{11}^{+} = \partial_y,\\
    ik_x = B_{21}^{+^\star}&\rightarrow& B_{21}^{+} = -ik_x,\\
    \gamma^2 = B_{32}^{+^\star}&\rightarrow& B_{32}^{+} = \gamma^2.
\end{eqnarray}
By arranging these results into matrix form we obtain the expression of the adjoint control matrix,
\begin{equation}
    B^+ = \begin{bmatrix}\partial_y &0 \\ -ik_x &0 \\0 &\gamma^2 \end{bmatrix}.
\end{equation}
Finally, computing the product $BB^+$ gives,
\begin{equation}
    BB^+ = \begin{bmatrix}
        \nabla^{-2}\partial_y &-ik_x\nabla^{-2} & 0\\
        0& 0&1
    \end{bmatrix}\begin{bmatrix}\partial_y &0 \\ -ik_x &0 \\0 &\gamma^2 \end{bmatrix} = \begin{bmatrix}
        1&0\\0&\gamma^2
    \end{bmatrix}.
\end{equation}

\bibliographystyle{unsrtnatabbrv}
\bibliography{references}

@article{podvin2012proper,
	title={{Proper orthogonal decomposition investigation of turbulent Rayleigh-B{\'e}nard convection in a rectangular cavity}},
	author={Podvin, B{\'e}reng{\`e}re and Sergent, Anne},
	journal={Phys. Fluids},
	volume={24},
	number={10},
	year={2012},
	publisher={AIP Publishing}
}

@article{cavalieri2022reduced,
  title={{Reduced-order Galerkin models of plane Couette flow}},
  author={Cavalieri, Andr{\'e} V. G. and Nogueira, Petr{\^o}nio A. S.},
  journal={Phys. Rev. Fluids},
  volume={7},
  number={10},
  pages={L102601},
  year={2022},
  publisher={APS}
}

@inproceedings{cavalieri2023non,
  title={{Non-linear Galerkin reduced-order models of a mixing layer}},
  author={Cavalieri, Andr{\'e} V. G.},
  booktitle={AIAA AVIATION 2023 Forum},
  pages={4483},
  year={2023}
}

@article{mccormack2024multi,
  title={{Multi-scale invariant solutions in plane Couette flow: a reduced-order model approach}},
  author={McCormack, Matthew and Cavalieri, Andr{\'e} V. G. and Hwang, Yongyun},
  journal={J. Fluid Mech.},
  volume={983},
  pages={A33},
  year={2024},
  publisher={Cambridge University Press}
}

@article{ilak2008modeling,
  title={{Modeling of transitional channel flow using balanced proper orthogonal decomposition}},
  author={Ilak, Milo{\v{s}} and Rowley, Clarence W.},
  journal={Phys. Fluids},
  volume={20},
  number={3},
  year={2008},
  publisher={AIP Publishing}
}

@article{jovanovic2005componentwise,
  title={{Componentwise energy amplification in channel flows}},
  author={Jovanovi{\'c}, Mihailo R. and Bamieh, Bassam},
  journal={J. Fluid Mech.},
  volume={534},
  pages={145--183},
  year={2005},
  publisher={Cambridge University Press}
}

@article{bagheri2009input,
  title={{Input-output analysis and control design applied to a linear model of spatially developing flows}},
  author={Bagheri, S. and Henningson, D. S. and H{\oe}pffner, J. and Schmid, Peter J.},
  journal={Appl. Mech. Rev.},
  volume={62},
  number={2},
  pages={020803},
  year={2009}
}

@article{waleffe1992nature,
  title={{The nature of triad interactions in homogeneous turbulence}},
  author={Waleffe, Fabian},
  journal={Phys. Fluids A},
  volume={4},
  number={2},
  pages={350--363},
  year={1992},
  publisher={American Institute of Physics}
}

@article{moore1973two,
  title={{Two-dimensional Rayleigh-B{\'e}nard convection}},
  author={Moore, D. R. and Weiss, N. O.},
  journal={J. Fluid Mech.},
  volume={58},
  number={2},
  pages={289--312},
  year={1973},
  publisher={Cambridge University Press}
}

@article{burns2020dedalus,
  title={{Dedalus: A flexible framework for numerical simulations with spectral methods}},
  author={Burns, Keaton J. and Vasil, Geoffrey M. and Oishi, Jeffrey S. and Lecoanet, Daniel and Brown, Benjamin P.},
  journal={Phys. Rev. Res.},
  volume={2},
  number={2},
  pages={023068},
  year={2020},
  publisher={APS}
}

@article{waleffe1997self,
  title={{On a self-sustaining process in shear flows}},
  author={Waleffe, Fabian},
  journal={Phys. Fluids},
  volume={9},
  number={4},
  pages={883--900},
  year={1997},
  publisher={American Institute of Physics}
}

@article{noack2003hierarchy,
  title={{A hierarchy of low-dimensional models for the transient and post-transient cylinder wake}},
  author={Noack, Bernd R. and Afanasiev, Konstantin and Morzy{\'n}ski, Marek and Tadmor, Gilead and Thiele, Frank},
  journal={J. Fluid Mech.},
  volume={497},
  pages={335--363},
  year={2003},
  publisher={Cambridge University Press}
}

@article{paul2011bifurcations,
  title={{Bifurcations and chaos in large-Prandtl number Rayleigh--B{\'e}nard convection}},
  author={Paul, Supriyo and Wahi, Pankaj and Verma, Mahendra K.},
  journal={Int. J. Non-Linear Mech.},
  volume={46},
  number={5},
  pages={772--781},
  year={2011},
  publisher={Elsevier}
}

@article{paul2012bifurcation,
  title={{Bifurcation analysis of the flow patterns in two-dimensional Rayleigh--B{\'e}nard convection}},
  author={Paul, Supriyo and Verma, Mahendra K. and Wahi, Pankaj and Reddy, Sandeep K. and Kumar, Krishna},
  journal={Int. J. Bifurcation Chaos},
  volume={22},
  number={05},
  pages={1230018},
  year={2012},
  publisher={World Scientific}
}

@article{ahmed2021resolvent,
  title={{Resolvent analysis of stratification effects on wall-bounded shear flows}},
  author={Ahmed, M. A. and Bae, H. J. and Thompson, A. F. and McKeon, B. J.},
  journal={Phys. Rev. Fluids},
  volume={6},
  number={8},
  pages={084804},
  year={2021},
  publisher={APS}
}

@article{lorenz1955available,
  title={{Available potential energy and the maintenance of the general circulation}},
  author={Lorenz, Edward N.},
  journal={Tellus},
  volume={7},
  number={2},
  pages={157--167},
  year={1955},
  publisher={Taylor \& Francis}
}

@article{saltzman1962finite,
  title={{Finite amplitude free convection as an initial value problem—I}},
  author={Saltzman, Barry},
  journal={J. Atmos. Sci.},
  volume={19},
  number={4},
  pages={329--341},
  year={1962}
}

@article{winters1995available,
  title={{Available potential energy and mixing in density-stratified fluids}},
  author={Winters, Kraig B. and Lombard, Peter N. and Riley, James J. and D'Asaro, Eric A.},
  journal={J. Fluid Mech.},
  volume={289},
  pages={115--128},
  year={1995},
  publisher={Cambridge University Press}
}

@article{hughes2013available,
  title={{Available potential energy in Rayleigh--B{\'e}nard convection}},
  author={Hughes, Graham O. and Gayen, Bishakhdatta and Griffiths, Ross W.},
  journal={J. Fluid Mech.},
  volume={729},
  pages={R3},
  year={2013},
  publisher={Cambridge University Press}
}

@article{mclaughlin1982transition,
  title={{Transition from periodic to chaotic thermal convection}},
  author={McLaughlin, John B. and Orszag, Steven A.},
  journal={J. Fluid Mech.},
  volume={122},
  pages={123--142},
  year={1982},
  publisher={Cambridge University Press}
}

@article{curry1984order,
  title={{Order and disorder in two-and three-dimensional B{\'e}nard convection}},
  author={Curry, James H. and Herring, Jackson R. and Loncaric, Josip and Orszag, Steven A.},
  journal={J. Fluid Mech.},
  volume={147},
  pages={1--38},
  year={1984},
  publisher={Cambridge University Press}
}

@article{yahata1982transition,
  title={{Transition to turbulence in the Rayleigh-B{\'e}nard convection}},
  author={Yahata, Hideo},
  journal={Prog. Theor. Phys.},
  volume={68},
  number={4},
  pages={1070--1081},
  year={1982},
  publisher={Oxford University Press}
}

@article{yahata1983period,
  title={{Period-doubling cascade in the Rayleigh-B{\'e}nard convection}},
  author={Yahata, Hideo},
  journal={Prog. Theor. Phys.},
  volume={69},
  number={6},
  pages={1802--1805},
  year={1983},
  publisher={Oxford University Press}
}

@book{getling1998rayleigh,
  title={{Rayleigh-B{\'e}nard convection: structures and dynamics}},
  author={Getling, Alexander V.},
  volume={11},
  year={1998},
  publisher={World Scientific}
}

@article{hindmarsh1983odepack,
  title={{ODEPACK, a systemized collection of ODE solvers}},
  author={Hindmarsh, Alan C.},
  journal={Scientific Computing},
  year={1983},
  publisher={North-Holland}
}

@article{petzold1983automatic,
  title={{Automatic selection of methods for solving stiff and nonstiff systems of ordinary differential equations}},
  author={Petzold, Linda},
  journal={SIAM J. Sci. Stat. Comput.},
  volume={4},
  number={1},
  pages={136--148},
  year={1983},
  publisher={SIAM}
}

@article{puigjaner2011steady,
  title={{From steady solutions to chaotic flows in a Rayleigh--B{\'e}nard problem at moderate Rayleigh numbers}},
  author={Puigjaner, D. and Herrero, J. and Sim{\'o}, C. and Giralt, F.},
  journal={Physica D},
  volume={240},
  number={11},
  pages={920--934},
  year={2011},
  publisher={Elsevier}
}

@article{pallares1999flow,
  title={{Flow transitions in laminar Rayleigh--B{\'e}nard convection in a cubical cavity at moderate Rayleigh numbers}},
  author={Pallares, Jordi and Grau, Francesc Xavier and Giralt, Francesc},
  journal={Int. J. Heat Mass Transfer},
  volume={42},
  number={4},
  pages={753--769},
  year={1999},
  publisher={Elsevier}
}

@article{pallares1996natural,
  title={{Natural convection in a cubical cavity heated from below at low Rayleigh numbers}},
  author={Pallares, J. and Cuesta, I. and Grau, F. X. and Giralt, Francesc},
  journal={Int. J. Heat Mass Transfer},
  volume={39},
  number={15},
  pages={3233--3247},
  year={1996},
  publisher={Elsevier}
}

@article{puigjaner2004stability,
  title={{Stability analysis of the flow in a cubical cavity heated from below}},
  author={Puigjaner, D. and Herrero, J. and Giralt, Francesc and Sim{\'o}, C.},
  journal={Phys. Fluids},
  volume={16},
  number={10},
  pages={3639--3655},
  year={2004},
  publisher={AIP Publishing}
}

@article{puigjaner2008bifurcation,
  title={{Bifurcation analysis of steady Rayleigh--B{\'e}nard convection in a cubical cavity with conducting sidewalls}},
  author={Puigjaner, Dolors and Herrero, Joan and Simo, Carles and Giralt, Francesc},
  journal={J. Fluid Mech.},
  volume={598},
  pages={393--427},
  year={2008},
  publisher={Cambridge University Press}
}

@article{puigjaner2006bifurcation,
  title={{Bifurcation analysis of multiple steady flow patterns for Rayleigh-B{\'e}nard convection in a cubical cavity at Pr= 130}},
  author={Puigjaner, D. and Herrero, J. and Giralt, Francesc and Sim{\'o}, C.},
  journal={Phys. Rev. E},
  volume={73},
  number={4},
  pages={046304},
  year={2006},
  publisher={APS}
}

@article{winchester2022onset,
  title={{The onset of zonal modes in two-dimensional Rayleigh--B{\'e}nard convection}},
  author={Winchester, Philip and Howell, Peter D. and Dallas, Vassilios},
  journal={J. Fluid Mech.},
  volume={939},
  pages={A8},
  year={2022},
  publisher={Cambridge University Press}
}

@article{wang2020zonal,
  title={{From zonal flow to convection rolls in Rayleigh--B{\'e}nard convection with free-slip plates}},
  author={Wang, Qi and Chong, Kai Leong and Stevens, Richard J. A. M. and Verzicco, Roberto and Lohse, Detlef},
  journal={J. Fluid Mech.},
  volume={905},
  pages={A21},
  year={2020},
  publisher={Cambridge University Press}
}

@article{munch2008galerkin,
  title={{Galerkin method for feedback controlled Rayleigh--B{\'e}nard convection}},
  author={M{\"u}nch, Andreas and Wagner, Barbara},
  journal={Nonlinearity},
  volume={21},
  number={11},
  pages={2625},
  year={2008},
  publisher={IOP Publishing}
}

@article{manneville1983two,
  title={{A two-dimensional model for three-dimensional convective patterns in wide containers}},
  author={Manneville, P.},
  journal={J. Phys.},
  volume={44},
  number={7},
  pages={759--765},
  year={1983},
  publisher={Soci{\'e}t{\'e} Fran{\c{c}}aise de Physique}
}

@article{rayleigh1916lix,
  title={{LIX. On convection currents in a horizontal layer of fluid, when the higher temperature is on the under side}},
  author={Rayleigh, Lord},
  journal={Phil. Mag.},
  volume={32},
  number={192},
  pages={529--546},
  year={1916},
  publisher={Taylor \& Francis}
}

@book{benard1901tourbillons,
  title={{Les tourbillons cellulaires dans une nappe liquide propageant de la chaleur par convection, en r{\'e}gime permanent}},
  author={B{\'e}nard, Henri},
  year={1901},
  publisher={Gauthier-Villars}
}

@article{busse1978non,
  title={{Non-linear properties of thermal convection}},
  author={Busse, Friedrich H.},
  journal={Rep. Prog. Phys.},
  volume={41},
  number={12},
  pages={1929},
  year={1978},
  publisher={IOP Publishing}
}

@article{busse1967stability,
  title={{The stability of finite amplitude cellular convection and its relation to an extremum principle}},
  author={Busse, Friedrich H.},
  journal={J. Fluid Mech.},
  volume={30},
  number={4},
  pages={625--649},
  year={1967},
  publisher={Cambridge University Press}
}

@article{busse2005transition,
  title={{Transition to turbulence in Rayleigh-Be{\'e}nard convection}},
  author={Busse, Friedrich H.},
  journal={Hydrodynamic instabilities and the transition to turbulence},
  pages={97--137},
  year={2005},
  publisher={Springer}
}

@article{bodenschatz2000recent,
  title={{Recent developments in Rayleigh-B{\'e}nard convection}},
  author={Bodenschatz, Eberhard and Pesch, Werner and Ahlers, Guenter},
  journal={Annu. Rev. Fluid Mech.},
  volume={32},
  number={1},
  pages={709--778},
  year={2000},
  publisher={Annual Reviews}
}

@book{chandrasekhar2013hydrodynamic,
  title={{Hydrodynamic and hydromagnetic stability}},
  author={Chandrasekhar, Subrahmanyan},
  year={2013},
  publisher={Courier Corporation}
}

@incollection{manneville2006rayleigh,
  title={{Rayleigh-B{\'e}nard convection: thirty years of experimental, theoretical, and modeling work}},
  author={Manneville, Paul},
  booktitle={Dynamics of spatio-temporal cellular structures: Henri B{\'e}nard centenary review},
  pages={41--65},
  year={2006},
  publisher={Springer}
}

@article{lorenz1963deterministic,
  title={{Deterministic nonperiodic flow}},
  author={Lorenz, Edward N.},
  journal={J. Atmos. Sci.},
  volume={20},
  number={2},
  pages={130--141},
  year={1963}
}

@article{grebogi1983crises,
  title={{Crises, sudden changes in chaotic attractors, and transient chaos}},
  author={Grebogi, Celso and Ott, Edward and Yorke, James A.},
  journal={Physica D},
  volume={7},
  number={1-3},
  pages={181--200},
  year={1983},
  publisher={Elsevier}
}

@article{zienicke1998bifurcations,
  title={{Bifurcations in two-dimensional Rayleigh-B{\'e}nard convection}},
  author={Zienicke, Egbert and Seehafer, Norbert and Feudel, Fred},
  journal={Phys. Rev. E},
  volume={57},
  number={1},
  pages={428},
  year={1998},
  publisher={APS}
}

@article{cavalieri2021structure,
  title={{Structure interactions in a reduced-order model for wall-bounded turbulence}},
  author={Cavalieri, Andr{\'e} V. G.},
  journal={Phys. Rev. Fluids},
  volume={6},
  number={3},
  pages={034610},
  year={2021},
  publisher={APS}
}

@article{muller1988convection,
  title={{Convection and inhomogeneities in crystal growth from the melt}},
  author={Muller, Georg},
  journal={Crystals},
  volume={12},
  pages={1--32},
  year={1988},
  publisher={Springer}
}

@article{alloui2018control,
  title={{Control of Rayleigh-B{\'e}nard convection in a fluid layer with internal heat generation}},
  author={Alloui, Z. and Alloui, Y. and Vasseur, Patrick},
  journal={Micrograv. Sci. Technol.},
  volume={30},
  pages={885--897},
  year={2018},
  publisher={Springer}
}

@article{gunzburger2002controlling,
  title={{Controlling transport phenomena in the Czochralski crystal growth process}},
  author={Gunzburger, M. and Ozugurlu, Ersin and Turner, J. and Zhang, H.},
  journal={J. Cryst. Growth},
  volume={234},
  number={1},
  pages={47--62},
  year={2002},
  publisher={Elsevier}
}

@article{cavalieri2022transition,
  title={{Transition to chaos in a reduced-order model of a shear layer}},
  author={Cavalieri, Andr{\'e} V. G. and Rempel, Erico L. and Nogueira, Petr{\^o}nio A. S.},
  journal={J. Fluid Mech.},
  volume={932},
  pages={A43},
  year={2022},
  publisher={Cambridge University Press}
}

@article{kashinath2014nonlinear,
  title={{Nonlinear self-excited thermoacoustic oscillations of a ducted premixed flame: bifurcations and routes to chaos}},
  author={Kashinath, Karthik and Waugh, Iain C. and Juniper, Matthew P.},
  journal={J. Fluid Mech.},
  volume={761},
  pages={399--430},
  year={2014},
  publisher={Cambridge University Press}
}

@article{geist1990comparison,
  title={{Comparison of different methods for computing Lyapunov exponents}},
  author={Geist, Karlheinz and Parlitz, Ulrich and Lauterborn, Werner},
  journal={Prog. Theor. Phys.},
  volume={83},
  number={5},
  pages={875--893},
  year={1990},
  publisher={Oxford University Press}
}

@article{jolly2011numerical,
  title={{Numerical Techniques for Approximating Lyapunov Exponents and Their Implementation}},
  author={Jolly, Michael S. and Van Vleck, Erik S.},
  journal={J. Comput. Nonlinear Dyn.},
  volume={6},
  pages={011003--1},
  year={2011}
}

@article{jimenez1991minimal,
  title={{The minimal flow unit in near-wall turbulence}},
  author={Jim{\'e}nez, Javier and Moin, Parviz},
  journal={J. Fluid Mech.},
  volume={225},
  pages={213--240},
  year={1991},
  publisher={Cambridge University Press}
}

@book{brunton2022data,
  title={{Data-driven science and engineering: Machine learning, dynamical systems, and control}},
  author={Brunton, Steven L. and Kutz, J. Nathan},
  year={2022},
  publisher={Cambridge University Press}
}

@article{rowley2017model,
  title={{Model reduction for flow analysis and control}},
  author={Rowley, Clarence W. and Dawson, Scott T. M.},
  journal={Annu. Rev. Fluid Mech.},
  volume={49},
  number={1},
  pages={387--417},
  year={2017},
  publisher={Annual Reviews}
}

@article{van2013comparison,
  title={{Comparison between two-and three-dimensional Rayleigh--B{\'e}nard convection}},
  author={Van Der Poel, Erwin P. and Stevens, Richard J. A. M. and Lohse, Detlef},
  journal={J. Fluid Mech.},
  volume={736},
  pages={177--194},
  year={2013},
  publisher={Cambridge University Press}
}

@article{johnston2009comparison,
  title={{Comparison of turbulent thermal convection between conditions of constant temperature and constant flux}},
  author={Johnston, Hans and Doering, Charles R.},
  journal={Phys. Rev. Lett.},
  volume={102},
  number={6},
  pages={064501},
  year={2009}
}

@article{turton2015prediction,
  title={{Prediction of frequencies in thermosolutal convection from mean flows}},
  author={Turton, Sam E. and Tuckerman, Laurette S. and Barkley, Dwight},
  journal={Phys. Rev. E},
  volume={91},
  number={4},
  pages={043009},
  year={2015},
  publisher={APS}
}

@book{anderson1999lapack,
  title={LAPACK users' guide},
  author={Anderson, Edward and Bai, Zhaojun and Bischof, Christian and Blackford, L Susan and Demmel, James and Dongarra, Jack and Du Croz, Jeremy and Greenbaum, Anne and Hammarling, Sven and McKenney, Alan and others},
  year={1999},
  publisher={SIAM}
}

@article{soucasse2019proper,
  title={{Proper orthogonal decomposition analysis and modelling of large-scale flow reorientations in a cubic Rayleigh--B{\'e}nard cell}},
  author={Soucasse, Laurent and Podvin, B{\'e}reng{\`e}re and Rivi{\`e}re, Philippe and Soufiani, Anouar},
  journal={J. Fluid Mech.},
  volume={881},
  pages={23--50},
  year={2019},
  publisher={Cambridge University Press}
}

@article{maia2025turbulence,
  title={{Turbulence suppression in plane Couette flow using reduced-order models}},
  author={Maia, Igor A and Cavalieri, Andr{\'e}},
  journal={J. Fluid Mech.},
  volume={1014},
  pages={A18},
  year={2025},
  publisher={Cambridge University Press}
}

@article{podvin2001low,
  title={Low-order models for the flow in a differentially heated cavity},
  author={Podvin, B{\'e}reng{\`e}re and Le Qu{\'e}r{\'e}, Patrick},
  journal={Phys. Fluids},
  volume={13},
  number={11},
  pages={3204--3214},
  year={2001},
  publisher={American Institute of Physics}
}

@article{howle1996comparison,
  title={{A comparison of the reduced Galerkin and pseudo-spectral methods for simulation of steady Rayleigh-B{\'e}nard convection}},
  author={Howle, LE},
  journal={Int. J. Heat Mass Transf.},
  volume={39},
  number={12},
  pages={2401--2407},
  year={1996},
  publisher={Elsevier}
}

@book{trefethen2000spectral,
  title={Spectral methods in MATLAB},
  author={Trefethen, Lloyd N},
  year={2000},
  publisher={SIAM}
}

@article{liao2025convex,
  title={A Convex Optimization Approach to Compute Trapping Regions for Lossless Quadratic Systems},
  author={Liao, Shih-Chi and Leonid Heide, A and Hemati, Maziar S and Seiler, Peter J},
  journal={Int. J. Robust Nonlinear Control},
  volume={35},
  number={6},
  pages={2425--2436},
  year={2025},
  publisher={Wiley Online Library}
}

@article{ahmed2021closures,
  title={{On closures for reduced order models—A spectrum of first-principle to machine-learned avenues}},
  author={Ahmed, Shady E and Pawar, Suraj and San, Omer and Rasheed, Adil and Iliescu, Traian and Noack, Bernd R},
  journal={Phys. Fluids},
  volume={33},
  number={9},
  year={2021},
  publisher={AIP Publishing}
}

@article{flores2026state,
	title   = {{State estimation of {R}ayleigh--{B}\'enard convection with reduced-order models}},
	author  = {Flores-Montoya, Enrique and da Silva, Andr\'e F. C. and Cavalieri, Andr\'e V. G.},
	journal={arXiv preprint arXiv:2606.20511},
	year    = {2026}
}

@article{ascher1997implicit,
	title={Implicit-explicit Runge-Kutta methods for time-dependent partial differential equations},
	author={Ascher, Uri M and Ruuth, Steven J and Spiteri, Raymond J},
	journal={Applied Numerical Mathematics},
	volume={25},
	number={2-3},
	pages={151--167},
	year={1997},
	publisher={Elsevier}
}
\end{document}